\documentclass[10pt,a4paper,twoside]{article}
\usepackage[T1]{fontenc}
\usepackage[utf8]{inputenc}
\usepackage{geometry}
\usepackage{amsmath, amssymb}
\usepackage{braket}
\usepackage{quantikz} 
\usepackage{graphicx}
\usepackage{subcaption}
\usepackage{float}
\usepackage{booktabs}
\usepackage{array}
\usepackage{colortbl}
\usepackage{enumitem}
\setlist{nosep}
\usepackage{xcolor}
\usepackage{tikz}
\usetikzlibrary{
  arrows.meta,
  positioning,
  shapes.geometric,
  calc,
  shadows
}
\usepackage[hidelinks]{hyperref}
\usepackage{titling} 
\usepackage{fancyhdr}
\fancypagestyle{firstpage}{
    \fancyhf{}

    \fancyfoot[L]{\footnotesize 
        
    } 
}

\title{\textbf{Quantum Simulation of SPAD in the Space Radiation Environment}}
\author{
    Durgesh Tinker\textsuperscript{$\dagger$,1} and Kavita Lalwani\textsuperscript{*,1} \\[1ex]
    \normalsize \textit{\textsuperscript{1}Department of Physics, Malaviya National Institute of Technology Jaipur, India}\\
    \small 
    \textsuperscript{*} Corresponding author: \texttt{\textit{kavita.phy@mnit.ac.in}}\\
     \small
    \textsuperscript{$\dagger$} \texttt{\textit{2025rpy9085@mnit.ac.in}}
}
\date{} 
\begin{document}
\maketitle
\thispagestyle{firstpage}

\begin{abstract}
Single-Photon Avalanche Diodes (SPADs) are critical components of emerging quantum communication networks that detect single photons. They are placed in satellites for long-distance communication and are susceptible to radiation-induced displacement damage in space. This degrades SPAD performance parameters, including reduced efficiency, increased thermal dark counts, damage to the Si crystal, and increased afterpulsing rate. This paper simulates SPAD in a space radiation environment by introducing a quantum simulation framework, modeling the SPAD detector as a three-level quantum system, ground state ($|g\rangle$), excited state ($|e\rangle$) and a trap state ($|t\rangle$). Furthermore, to model the photon as a quantum system, second quantization and Fock-space truncation are used. The interaction between the photon-SPAD closed system is simulated using the Jaynes-Cummings model, and the open-system dynamics is governed by the Lindblad master equation and Qiskit's gate-based noise channels. The key characteristics, such as the efficiency, timing jitter, thermal dark counts, and afterpulsing \& radiation effects, are obtained using quantum simulation of SPAD. The efficiency of the SPAD is visualized through Rabi oscillations. Timing jitter is calculated as the FWHM of the detection probability curve, which is approximately 720\,ps. When photon dissipation is introduced, the Rabi oscillations show damping, and the peak-detection probability decreases, and timing jitter increases. The thermal dark count rates are simulated using the Arrhenius model with an activation energy $\Delta E = 0.40$\,eV over the temperatures $T = 173, 273, 303$\,K. The dark count rate is increased from from 2.22\,cps at 173\,K to $2.22 \times 10^5$\, cps at 303\,K over a temperature range of only 130\, K.  The afterpulsing is simulated using a Kraus channel. The power law is used to model the non-Markovian nature of afterpulsing. While the Markovian assumption overestimates the de-trapping ($\approx77\%$), the non-Markovian approach reveals that only $\approx50\%$ of traps are vacated by the end of the dead time ($\tau_0$). Which means the remaining 50\% of charge carriers in trap states may appear as false detections. This approach differs significantly from conventional TCAD simulations, which rely on semiclassical approximations that do not capture the discrete quantum statistics of single-photon interactions and the dynamics of trap states. 
\end{abstract}
\section{Introduction}
Single-Photon Avalanche Diodes (SPADs) \cite{hofbauer} are essentially quantum devices used primarily to detect single photons. SPADs play a crucial role in single-photon detection in many technologies, such as Quantum key distribution (QKD) and deep-space optical communication. It is fundamentally a semiconductor junction diode operated in reverse bias, working in the Geiger mode \cite{dsouza}. In this mode, the electric field is sufficiently high that a single charge carrier (electron), generated by the absorption of an incoming photon, can gain enough kinetic energy to further ionize the lattice atoms and trigger a self-sustaining avalanche of secondary carriers in the depletion region. This mechanism allows the device to convert a single quantum of light into a macroscopic current pulse. This unique operation provides high photon detection efficiency, low timing jitter (on the order of picoseconds \cite{InGaAs_SPAD}), and high gain, making SPADs very important for high-energy and space physics applications \cite{Hadfield, Helleboid}.

The space radiation environment \cite{Taylor} primarily comes from three different sources: Galactic cosmic rays (GCRs), Solar energetic particles (SEP), and trapped particles. The GCR originates from beyond our solar system, and its spectrum consists of protons of about 85-87 \%, 10-12 \% of alpha particles, and the remaining are heavy ions \cite{Taylor}.  The SEP is released during solar flares and Coronal mass ejections (CMEs).  The SEP contains mainly protons of about 90-95 \%, and the remaining are heavy nuclei, which are of more concern during solar maxima \cite{Taylor}. In Earth’s magnetic field, low-energy particles get trapped and create the Van Allen belts.

Satellites orbiting in Low Earth orbit (LEO) or in deep space are continuously exposed to high-energy charged particles (protons and electrons) trapped in the Van Allen belts. These charged-particle radiations cause damage to spacecraft, detectors (for example, SPADs), and their electronic components, posing a risk to space missions \cite{disdam}.  When a high-energy particle (example proton) collides with Si-SPAD, it displaces the atoms in Si lattice, forming stable interstitial-vacancy defects, which introduce mid-gap energy states, or traps, within the semiconductor bandgap \cite{Taylor, disdam}.  These traps act as generation \& recombination centers, which lead to degradations of the SPAD detector over mission lifetimes \cite{anisimova}, such as reduction in photon detection efficiency, increasing the thermal dark count rate, increasing afterpulse probabilities, altering the timing jitter, and enhancing the noise level, which are important characteristics of SPADs \cite{Helleboid}.

Here we summarize the previous work, which has been performed on SPAD theoretically, experimentally, as well as using simulations along with appropriate references, which will help us to understand the current status of SPAD detector work and the need of this work (Quantum Simulations of SPAD for space applications). The radiation hardness of SPADs is currently modeled using Technology Computer-Aided Design (TCAD) simulation tools \cite{TCAD}.  It is a powerful simulation tool for modeling the device physics of the SPAD detector and is extremely helpful for bulk modeling \cite{TCAD}. However, these simulations rely on continuous-fluid approximations that struggle to capture the inherently discrete and stochastic nature of single-photon interactions at the device level \cite{TCAD drawback}. These classical methods often treat trap dynamics and photon absorption as averaged statistical processes \cite{wei2024, wang2025} and do not address the fundamental quantum statistics of single-photon interactions or the quantum open-system dynamics intrinsic to SPAD operation, assuming light as an EM wave. In SPAD detectors, where a single photon interacts with the detector material and the interaction is essentially quantum in nature, this limitation becomes critical when modeling quantum observables such as detector efficiency, afterpulsing, thermal dark counts, and radiation-induced noise for space applications.

In the previous simulation effort by Wei et al. \cite{wei2024}, they conducted a detailed simulation and theoretical analysis of the measurement errors induced by afterpulsing in InGaAs SPADs. They used a mathematical power-law equation to describe the experimental results. Furthermore, the influence of afterpulse probability and dead time on the system’s average count rate was also analyzed. This simulation study provided insights into the importance of the power law in trapped-charge-carrier detrapping and its effects on afterpulsing probability. Further, Wang et al. \cite{wang2025} conducted a simulation study using non-Markovian afterpulsing into a closed-form symbol-error-rate model for free-space optical links, treating the power-law behavior as an empirical input to a classical receiver model. Both previous simulation works \cite{wei2024, wang2025} confirm that non-Markovian, power-law afterpulsing statistics outperform Markovian assumptions for real SPAD devices. This non-Markovian assumption differs significantly from the afterpulsing model by DSouza \cite{dsouza}, which assumes that all traps release at the same time with a constant rate, i.e., an exponential Markovian decay model, and that the afterpulsing effect is complete within a few microseconds after a dead time. 

In addition to the simulation studies, Anisimova et al. \cite{anisimova} conducted experiments on various avalanche photodiodes under a two-year-equivalent radiation exposure. They tested nine groups of samples at nominal radiation fluences ranging from 108 to 1010 $protons/cm^2$, corresponding to equivalent in-orbit exposures of 0.6, 6, 12, and 24 months, with a proton energy of 106 MeV. The effects of radiation damage on dark counts and afterpulsing were analyzed, and their mitigation techniques were discussed. They found that all detectors irradiated with protons showed a significant increase in dark counts, reaching saturation. They also reported an increase in afterpulsing probability for some SPADs of their test group. Further, another experimental study by Campajola et al. \cite{IEEE} also performed various displacement-damage dose experiments, and it was found that DCR followed the Arrhenius law \cite{IEEE}, with an activation energy of 0.4 eV. Both experimental studies \cite{anisimova, IEEE} reported a significant drop in DCR after annealing. It was observed that radiation-induced damages, i.e., displacement of Si atoms from their ground states to trap states, which is known as Non-ionizing energy loss (NIEL), increase almost linearly with increase in fluence, which is fundamentally expressed by $N_{trap} = \alpha \phi$ \cite{IEEE}. where $\phi$ is fluence and $\alpha$ is the rate of radiation damage. Furthermore, as the Afterpulsing probability is proportional to the trap population, afterpulsing should also increase almost linearly. Although this assumption failed completely in practical cases \cite{anisimova}, as afterpulsing probabilities were observed to be nonlinear for most SPADs in Anisimova's experimental studies, only one type of APD showed almost linear behavior. Hence, the first-order approximation indicates that the afterpulsing probability increases almost linearly with the external fluence.

However, the approaches by Wei (2024) and Wang (2025) \cite{wei2024, wang2025}  use classical statistical models to model the temporal evolution of afterpulsing probability. They do not model the trap-detrapping process, which is important for afterpulsing analysis. Wei \cite{wei2024} does not model the microscopic trap dynamics, nor connect trap physics to a quantum channel description. Wang's \cite{wang2025} non-Markovian model was a phenomenological statistical description embedded in a communication-system performance analysis and it does not provide a time-domain quantum-dynamical simulation of trap occupancy and release. Furthermore, radiation damage is not considered in either approach \cite{wei2024, wang2025} for the afterpulsing study. They use simplified analytical corrections, e.g., Poisson statistics for the dead-time and count-rate dependence. Furthermore, the experimental studies \cite{anisimova, IEEE} report the average noise after radiation damage, completely ignoring the quantum behavior of how individual charges are trapped and released. Physical hardware tests cannot separate the true physics of delayed charge release from the effects of the detector's electronic dead time. Existing classical models cannot dynamically link single-photon optical interactions with non-Markovian trap memory, necessitating a unified quantum simulation. Campajola et al. \cite{IEEE} evaluated radiation-induced trap states using semiclassical, time-averaged Shockley-Read-Hall approximations. This bulk approach is fundamentally unsuitable for capturing the non-Markovian afterpulsing effects. Furthermore, Anisimova \cite{anisimova} explicitly notes that the long dead time of their quenching circuit suppressed the observation of delayed afterpulsing. Hence, the true picture of afterpulsing can not be captured.
 
There is a need to develop a quantum simulation framework \cite{nielsen} to model the quantum-level interactions of the SPAD detector, expressing key characteristics as quantum-state transition probabilities to capture the inherent quantum nature of the photon-SPAD interaction. Hence, in this work, the SPAD detector is modeled as a three-level quantum system \cite{nath} with the energy levels $\ket{g}, \ket{t}\ \& \ket{e}$ \cite{nath}. Furthermore, the photon system is quantized. This approach differs significantly from the previous quantum simulation approach, which modeled SPAD as a two-level system (TLS) \cite{DTINKER, tinker2}. This TLS model alone is insufficient to obtain afterpulsing effects as it lacks the necessary structure, the trap level, to model trapping and subsequent release. This three-level system enables simulation of key characteristics, including SPAD efficiency, timing jitter, thermal dark counts, and afterpulsing. This work introduces a quantum simulation framework that leverages the formalism of open quantum systems \cite{OQS, Lidar2019} to model SPAD dynamics in a radiation environment. Furthermore, the interaction between a photon and a SPAD is described by the Jaynes-Cummings model \cite{jcmodel}, treating light as a quantum harmonic oscillator and truncating the photon Fock space. The Lindblad master equation \cite{GKLS, GKLS1, GKLS2, manzano}, Qiskit noise models \cite{qiskit, qn}, and  Kraus operators \cite{kraus} are used to simulate the SPAD detector in a space radiation environment. The key performance metrics, including efficiency, thermal dark counts, and afterpulsing, are then calculated from the transition probabilities between three SPAD levels and their expectation values. The motivation for this work is not to compete with classical TCAD simulations or the experimental approaches, but instead to develop a new framework for SPAD detectors based on Quantum simulations.

The paper is structured as follows. In Section 2, the theoretical framework and Mathematical formulation are described. This is followed by section 3, which describes the quantum simulation framework, including a simulation flowchart and the parameters used. In section 3, all simulated characteristics, such as efficiency, thermal dark counts, timing jitter, and afterpulsing, are presented.

\section{Theoretical framework and Mathematical formulation}
In this section, the modeling of the Photon and SPAD as the quantum system is described. The methodology is divided into 3 steps. In step one, the complex many-body physics of SPAD is reduced to a treatable quantum model. The mapping of the SPAD detcteor to a finite, discrete $N$-level open quantum system is justified by the specific operational mechanics of Geiger-mode detection and the nature of radiation-induced defects. The Giger mode exhibits a binary response means whenever a photon triggers an avalanche, the detector becomes blind to subsequent photons, effectively neglecting them \cite{Helleboid}. Therefore, to analyze the SPAD detector, the states $|g\rangle$ (armed/ground) and $|e\rangle$ (avalanche/excited) can represent the collective states of the SPAD, in place of simulating the vast number of Si atoms ($>10^{23}$ cm$^{-3}$), which is a tough task using the present NISQ era quantum computers. An effective starting point for quantifying the SPAD system is the Two-Level System (TLS) approximation, which captures the essential quantum event underlying photon detection \cite{jcmodel}, which is the absorption of a photon and the subsequent generation of an electron–hole pair. However, the TLS model is insufficient to analyze the trap states and afterpulsing effects \cite{DTINKER, tinker2}. 

To incorporate afterpulsing into quantum simulation, an additional energy level corresponding to a trap state is introduced. Consequently, the SPAD detector is modeled as a three-level system with energy levels corresponding to ground state ($\ket{g})$, trap state ($\ket{t}$) and excited state ($\ket{e}$), which gives an advantage over the previous TLS approach \cite{DTINKER, tinker2}. In step two, the photon state truncation \cite{tudorovskaya} is described. The Fock-state formulation is used to describe the number of photons in the quantum state, accounting for single-photon interactions with the SPAD. Standard Binary Mapping (SBE) \cite{SBE} is used to describe classical Hamiltonians obtained in steps one \& two, in terms of the two-qubit computational basis \cite{nielsen, qcg}. Furthermore, in step three, the interaction between the three-level SPAD model and a single photon is described using the Jaynes–Cummings Hamiltonian \cite{nielsen, jcmodel} and the system Hamiltonian is derived, enabling its implementation on a quantum computer using Qiskit. The flowchart summarizing the theoretical framework is shown below in figure~\ref{fig:photon_spad_modeling}.
\par\vspace{6pt}

\begin{figure}[H]
    \centering
    \begin{tikzpicture}[
        node distance=0.6cm, 
        box/.style={
            rectangle, 
            draw=blue!70!black, 
            fill=white, 
            thick, 
            rounded corners=4pt, 
            text width=8cm, 
            align=center, 
            inner sep=10pt,
            font=\small
        },
        arrow/.style={thick, ->, >=stealth, draw=black!70}
    ]

    \node (b1) [box] {
        \textbf{Step 1: Theoretical Modelling of SPAD}\\
      \textit{ The SPAD is modeled as a three-level quantum system, $\ket{g},\ \ket{t},\&\ 
       \ket{e}$, which is subsequently mapped to qubit basis states $\mathcal{B}$ using Standard Binary Encoding (SBE) \cite{SBE}} \\ 
    };

    \node (b2) [box, below=of b1] {
        \textbf{Step 2: Photon Field Quantization and State Truncation}\\
        \textit{The Photon Field is Quantization and truncated \cite{tudorovskaya} to have a maximum number of photons $N_{max}$ in the state. Furthermore, SBE is applied to represent the quantum Hamiltonian in the qubit basis states.}
    };


     \node (b3) [box, below=of b2] {
        \textbf{Step 3: Construction of system Hamiltonian}\\
       \textit{Jaynes-Cummings (JC) interaction is applied to model the interaction between Photon-SPAD quantum systems.}
    };

    \draw [arrow] (b1.south) -- (b2.north);
    \draw [arrow] (b2.south) -- (b3.north);

    \end{tikzpicture}
    \caption{Workflow for the modeling of the Photon and SPAD system.}
    \label{fig:photon_spad_modeling}
\end{figure}
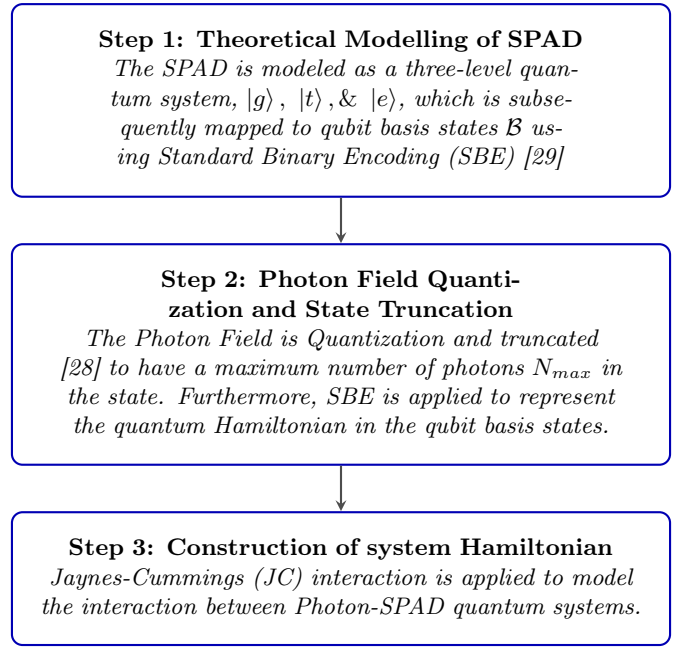

\subsection{Theoretical Modelling of SPAD} \label{SPADSBE}
\begin{figure}[H] \label{threelevel}
    \centering
    \begin{tikzpicture}
        \draw[thick] (0,0) -- (3,0) node[right]{$|g\rangle$ (Ground)};
        \draw[thick] (0,3) -- (3,3) node[right]{$|e\rangle$ (Avalanche)};
        \draw[thick, red] (1,2) -- (2,2) node[right]{$|t\rangle$ (Trap)};

        \draw[<->] (0.5,0) -- (0.5,3)
            node[midway, left]{Photon};
        \draw[->, dashed] (1.5,3) -- (1.5,2)
            node[midway, right, font=\tiny]{Trap};
        \draw[->, dashed] (1.2,2) -- (1.2,3)
            node[midway, left, font=\tiny]{Release};
    \end{tikzpicture}
    \caption{Three-level SPAD model showing ground, avalanche, and trap states with photon excitation and trapping dynamics}
    \label{fig:three_level_spad}
\end{figure}
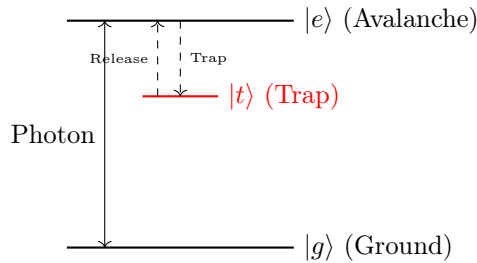

To understand the quantum-state transitions of the SPAD, it is modeled as a three-level quantum system with a ground state ($\ket{g}$), an excited state ($\ket{e}$), and a trap state ($\ket{t}$). These three states represent SPAD when it is armed (ready to detect), when an avalanche occurs, and when SPAD is at trap level, respectively. Figure.~(\ref{fig:three_level_spad}) shows the energy-level diagram of a three-level SPAD quantum system. The red-marked energy level illustrates the trap energy level. SPAD can undergo transitions not only between the ground and excited states, but also into trap energy
levels. The trapping mechanism occurs during the relaxation process from the excited state, where a charge carrier may be captured by a trap level and subsequently released at a later time. In this simulation, it is assumed that the transitions between the ground state $|g\rangle$ and the excited state $|e\rangle$ are governed by photon absorption, which is the ideal case of photon-SPAD interaction, whereas the transitions involving the trap state $|t\rangle$ (i.e., ground--trap and excited--trap transitions) arise due to interactions with the surrounding environmental noise of space radiation, primarily protons. The Hamiltonian $\text{H}_{\text{SPAD}}$ of this three-level SPAD system can be written with the help of projector operators and expressed in Eq.~(\ref{projector}).
\begin{equation} \label{projector}
        \text{H}_{\text{SPAD}}= E_g|g\rangle\langle g|+E_e|e\rangle\langle e|+E_t|t\rangle\langle t|
    \end{equation}

To simulate the SPAD detector with Quantum computing, the energy levels are mapped to qubit computational basis states \cite{nielsen, qcg}, which is known as \textbf{Standard Binary Encoding (SBE)} \cite{SBE}. Furthermore, any physical system with two levels can be represented as a Qubit \cite{nielsen, qiskit}. A qubit lies in a two-dimensional vector space of basis $\mathcal{B}=\{0,1\}$, as the SPAD detector is modeled as a three-level system, hence, a single qubit can not represent it. Hence, two qubits are used to represent a three-level SPAD quantum system whose basis vectors are written as a tensor product of their individual bases \cite{qcg}. The computational bases of the joint quantum system of three-level SPAD are $\mathcal{B}= \{00, 01, 10, 11\}$. The ground state $\ket{g}$ is mapped to $\ket{00}$, trap state $\ket{e}$ is mapped to $\ket{10}$ and excited state $\ket{e}$ is mapped to $\ket{01}$, while $\ket{11}$ remains unused or axillary state. The Hamiltonian of Eq.~(\ref{projector}) is hereby written in the form of these bases and expressed as Eq.~(\ref{BMH}).
\begin{equation} \label{BMH}
        \text{H}_{\text{SPAD}}= E_g|00\rangle\langle 00|+E_e|01\rangle\langle 01|+E_t|10\rangle\langle 10|
    \end{equation}

The matrix representations of two-qubit states follow the standard two-qubit computational basis written as a tensor product of one qubit basis matrices and expressed in Eq.~(\ref{matrixf}) \cite{nielsen, qcg}.
\begin{equation} \label{matrixf}
|00\rangle =
\begin{pmatrix}
1 \\ 0 \\ 0 \\ 0
\end{pmatrix}, \quad
|01\rangle =
\begin{pmatrix}
0 \\ 1 \\ 0 \\ 0
\end{pmatrix}, \quad
|10\rangle =
\begin{pmatrix}
0 \\ 0 \\ 1 \\ 0
\end{pmatrix}.
\end{equation}

Furthermore, putting values of Eq.~(\ref{matrixf}) in Eq.~(\ref{BMH}), the matrix form of SPAD Hamiltonian is obtained in the basis of two-qubit computational basis and expressed in Eq.~(\ref{ham}).
    \begin{equation} \label{ham}
        \text{H}_{\text{SPAD}}=\begin{pmatrix} E_g & 0 & 0 & 0 \\ 0 & E_e & 0 & 0 \\ 0 & 0 & E_t & 0 \\ 0 & 0 & 0 & 0 \end{pmatrix}
    \end{equation}

\subsection{Photon Field Quantization and State Truncation} \label{PFQ}

To simulate the single-photon-SPAD interaction, the electromagnetic field is quantized and then truncated. Following the quantization procedure of the photon field described in reference \cite{jjs, sakurai}, the single mode of the electromagnetic field is described as a quantum harmonic oscillator. The Hamiltonian for this photon field is expressed as Eq.~(\ref{jjs}).
\begin{equation} \label{jjs}
    \text{H}_{\mathrm{ph}} = \hbar \Omega \hat{a}^\dagger \hat{a}
\end{equation}
where $\Omega$ is the frequency of the optical mode, and $\hat{a}^\dagger$ and $\hat{a}$ are the bosonic creation and annihilation operators satisfying the canonical commutation relation $[\hat{a}, \hat{a}^\dagger] = 1$. The eigenstates of $\text{H}_{\mathrm{ph}}$ form an infinite-dimensional Fock space $\text{H}_{\mathrm{ph}} = \text{span}\{|n\rangle\}_{n=0}^{\infty}$, where the action of the operators is given by $a|n\rangle = \sqrt{n}|n-1\rangle$ and $a^\dagger|n\rangle = \sqrt{n+1}|n+1\rangle$ \cite{jjs}. The operator $N = a^\dagger a$ is the number operator, whose eigenvalue is the photon count $n$. A gate-based quantum computer operates on a finite Hilbert space $\mathcal{H}_{\mathrm{qubit}} \cong \mathbb{C}^{2^{n_q}}$, whose dimension increases polynomially with the number of qubits. Hence, the infinite-dimensional bosonic space is truncated to a cutoff, $N_{\mathrm{max}}$, restricting the occupation number of the photon mode \cite{HO}. Furthermore, the Hilbert space is now spanned by the basis states $\{\ket{0},\ket{1},\dots, \ket{N_{max}}\}$. The dimension of this truncated space for a single bosonic mode becomes $d = N_{max} + 1$. This truncation is mathematically formalized by introducing the projection operator $\mathcal{P}$, which maps the infinite space onto a finite $(N_{\mathrm{max}}+1)$-dimensional subspace. The projector operator is described in Eq.~(\ref{tmatrix}).
\begin{equation} \label{tmatrix}
    \mathcal{P} = \sum_{n=0}^{N_{\mathrm{max}}} |n\rangle \langle n|
\end{equation}
The continuous bosonic operators are thus mapped to their truncated finite-dimensional components \cite{tudorovskaya, HO}, $\tilde{a} = \mathcal{P} \hat{a} \mathcal{P}$ and $\tilde{a}^\dagger = \mathcal{P} \hat{a}^\dagger \mathcal{P}$. In the truncated Fock basis, the projected creation operator takes the explicit form described in Eq.~(\ref{transformation}).
\begin{equation} \label{transformation}
    \tilde{a}^\dagger = \sum_{n=0}^{N_{\mathrm{max}}-1} \sqrt{n+1} |n+1\rangle\langle n|
\end{equation}

In this simulation, the case of photon-number truncation with $N_{max}=3$ is considered, which serves as a general case for studying $N_{max}=1$ and $2$. For $N_{max}=3$, the matrix representation of $a^\dagger$ in the basis $\{\ket{00}, \ket{01}, \ket{10}, \ket{11}\}$ is expressed in Eq.~(\ref{creation}).
\begin{equation} \label{creation}
    a^\dagger = \begin{pmatrix} 0 & 0 & 0 & 0 \\ \sqrt{1} & 0 & 0 & 0 \\ 0 & \sqrt{2} & 0 & 0 \\ 0 & 0 & \sqrt{3} & 0 \end{pmatrix},\ \text{With}\ a=(a^\dagger)^T
\end{equation}

Finally, to execute this truncated Hamiltonian on a quantum simulator, the states are encoded into qubits. The Standard Binary Encoding (SBE) is utilize \cite{SBE}, which requires $n_q = \lceil \log_2(N_{\mathrm{max}}+1) \rceil$ qubits \cite{tudorovskaya} to represent the bosonic mode. The physical Fock state $|n\rangle$ is mapped to the computational basis $|b_{n_q-1} \dots b_0\rangle$ through the binary decomposition, as described in Eq.~(\ref{binary}).
\begin{equation} \label{binary}
    n = \sum_{j=0}^{n_q-1} b_j 2^j, \quad b_j \in \{0, 1\}
\end{equation}
By setting $N_{\mathrm{max}}=3$ ($n_q=2$), the binary mapping of photon occupation number is $\ket{0} \mapsto \ket{00},\ \ket{1} \mapsto \ket{01},\ \ket{2} \mapsto \ket{10},\ \ket{3} \mapsto \ket{11}$, where each of $\ket{0}, \ket{1}, \ket{2}, \ket{3}$ represents number of photons in the system. \cite{SBE}.
\subsection{Construction of system Hamiltonian}
The interaction between an incoming photon and the SPAD detector is simulated using the Jaynes-Cummings (JC) model \cite{jcmodel}. This theoretical model provides the simplest interaction between a single-mode electromagnetic field and a two-level system. The system dynamics are governed by the Hamiltonian given in Eq. (\ref{jcm}).
\begin{equation} \label{jcm}
    H_{\text{JCM}} = \text{H}_{\text{Photon}} + \text{H}_{\text{SPAD}} + \text{H}_{\text{Interaction}}
\end{equation} 
This model couples the two states of the SPAD detector such that the excitation number remains conserved. If the system has n number of photons, and SPAD is in the ground state, then the quantum state of the combined photon-SPAD system is $|n, g\rangle$. Furthermore, upon absorbing a single photon, it transitions to the excited state. In this case, the quantum state is written as $|n-1, e\rangle$. Mathematically, it is expressed as Eq. (\ref{jcref}).
\begin{equation} \label{jcref}
    |n, g\rangle \leftrightarrow  |n-1, e\rangle
\end{equation}
Eq. (\ref{jcref}) is used to simulate the binary response of a SPAD operating in Geiger mode, regardless of the total number of incident photons, which justifies the use of this model to describe a single photon-SPAD
interaction. This approach allows to move beyond classical simulations and treat the SPAD detector as a quantum system. The JC model is extended to account for a three-level SPAD quantum system interacting with a single-mode electromagnetic field (Photon) \cite{nath, zhu, 3level}. The interaction Hamiltonian for Photon-SPAD system under the Rotating Wave Approximation (RWA) \cite{swine} is given as,
\begin{equation} \label{rwa}
    H_{int}=\hbar g (a^\dagger \sigma_{ge} + a\sigma_{eg})
\end{equation}
Here, $g$ is the photon-SPAD coupling constant, $\sigma_{ge}$ and $\sigma_{eg}$ are atomic lowering and raising operators whose matrix representation in the two-qubit computational basis states for a three-level system is given by the following Eq.~(\ref{lowrise}),
\begin{equation} \label{lowrise}
\begin{aligned}
\sigma_{eg} &= |e\rangle\langle g|
      = |01\rangle\langle 00|
      =
      \begin{pmatrix}
          0 & 0 & 0 & 0 \\
          1 & 0 & 0 & 0 \\
          0 & 0 & 0 & 0 \\
          0 & 0 & 0 & 0
      \end{pmatrix} \\[6pt]
\sigma_{ge} &= |g\rangle\langle e|
      = |00\rangle\langle 01|
      =
      \begin{pmatrix}
          0 & 1 & 0 & 0 \\
          0 & 0 & 0 & 0 \\
          0 & 0 & 0 & 0 \\
          0 & 0 & 0 & 0
      \end{pmatrix} \\[6pt]
\sigma_{eg}|g\rangle &= |e\rangle \qquad & \sigma_{eg}|e\rangle &= 0 \\
\sigma_{ge}|e\rangle &= |g\rangle \qquad & \sigma_{ge}|g\rangle &= 0
\end{aligned}
\end{equation}

The full interaction Hamiltonian includes processes that do not conserve energy. The Rotating Wave Approximation (RWA) \cite{swine} is a simplification that retains only the energy-conserving terms. The term $a\sigma_{eg}$ describes the SPAD absorbing a photon to move to the excited state, while
$a^\dagger\sigma_{ge}$ describes the SPAD emitting a photon as it de-excites. These two terms represent a resonant
exchange of energy. The other two terms, $a\sigma_{ge}$ (simultaneous annihilation of a photon and de-excitation of the
SPAD) and $a^\dagger\sigma_{eg}$ (simultaneous creation of a photon and excitation of the atom), violate energy
conservation. The RWA consists of neglecting these terms. Furthermore, the interaction Hamiltonian in Eq.~(\ref{rwa}) has two terms, each of which is a sum of two tensor products. By performing the tensor products the interaction Hamiltonian is obtained as follows.

\begin{equation} \label{16matrix}
    H_{\mathrm{int}} = \hbar g \begin{pmatrix}
    \mathcal{O} & \sigma_{eg} & \mathcal{O} & \mathcal{O} \\
    \sigma_{ge} &\mathcal{O} & \sqrt{2}\sigma_{eg} & \mathcal{O} \\
    \mathcal{O} & \sqrt{2}\sigma_{ge} & \mathcal{O} & \sqrt{3}\sigma_{eg} \\
    \mathcal{O} & \mathcal{O} & \sqrt{3}\sigma_{ge} & \mathcal{O}
    \end{pmatrix}
\end{equation}
where $\mathcal{O}$ represents a $4 \times 4$ null matrix, while $\sigma_{eg}\ \&\ \sigma_{ge}$ are the $4\times 4$ matrices defined earlier in Eq.~(\ref{lowrise}). The interaction Hamiltonian is an order $16 \times 16$\footnote{In Appendix \ref{appA} the full $16 \times 16$ is derived.}.
Hence, the full Hamiltonian for the three-level SPAD-Photon JC interaction can be written Eq. (\ref{full})
\begin{equation} \label{full}
    H_{\text{JCM}}=\text{H}_{\text{SPAD}}+\hbar\Omega  a^\dagger a+\hbar g (a^\dagger \sigma_{ge} + a\sigma_{eg})
\end{equation}

\section{Quantum Simulation Framework}
\begin{figure}[H]
    \centering
    \begin{tikzpicture}[
        node distance=0.6cm, 
        box/.style={
            rectangle, 
            draw=blue!70!black,
            thick, 
            rounded corners=4pt, 
            text width=8cm, 
            align=center, 
            inner sep=10pt,
            font=\small
        },
        arrow/.style={thick, ->, >=stealth, draw=black!70}
    ]


    \node (b2) [box] {
        \textbf{Step 1: Pauli String Mapping}\\
        \textit{Map the JC Hamiltonian to Pauli strings.}
    };

    \node (b3) [box, below=of b2] {
        \textbf{Step 2: Quantum Circuit Construction}\\
        \textit{Build the gate-based quantum circuits.}
    };

    \node (b4) [box, below=of b3] {
        \textbf{Step 3: Time Evolution}\\
        \textit{Evolve the system for both closed and open dynamics.}
    };

    \node (b5) [box, below=of b4] {
        \textbf{Step 4: Simulating key characteristics such as efficiency, thermal dark count, and afterpulsing as quantum expectation values (Results)}\\
    };

  
    \draw [arrow] (b2.south) -- (b3.north);
    \draw [arrow] (b3.south) -- (b4.north);
    \draw [arrow] (b4.south) -- (b5.north);

    \end{tikzpicture}
    \caption{Detailed step-by-step workflow for Hamiltonian mapping and quantum simulation evolution.}
    \label{fig:sim_steps}
\end{figure}
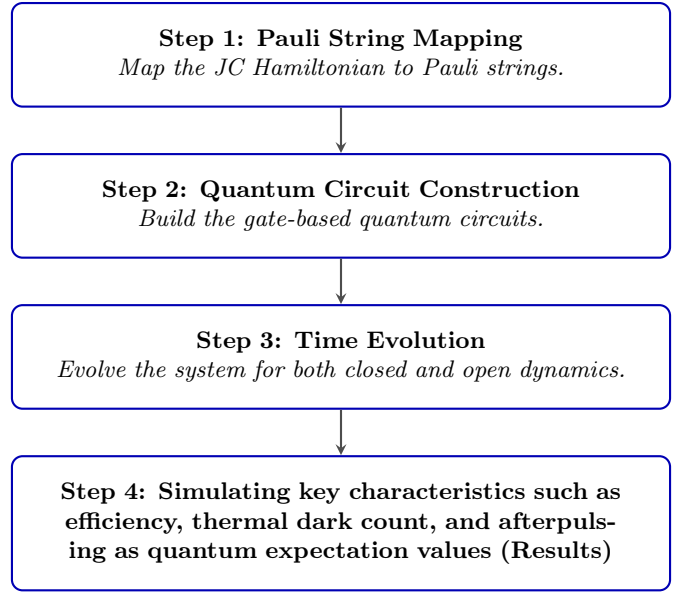

Quantum simulation accurately captures the quantum-level interactions between a photon and an SPAD in a space radiation environment. IBM Quantum's Qiskit SDK \cite{qiskit} is used to simulate the SPAD characteristics. The Quantum Simulation framework comprises 4 steps. In step one, the JC Hamiltonian of Eq.~(\ref{full}) is mapped to Pauli strings \cite{fondana}, which enables its
implementation on a quantum simulator \cite{aer, lloyd1996}. Qiskit's \texttt{SparsePauliOp} \cite{sp} is used to convert the JC Hamiltonian into Pauli strings. In step two, the quantum circuit is constructed according to the given characteristics, using qiskit's \texttt{QuantumCircuit} \cite{qc}. These circuits are then executed to simulate the system's time evolution, accounting for both ideal (closed system) and noisy (open system) environments, using Trotterization \cite{suzuki1991, trotterQ}, Lindblad master equation \cite{GKLS}, and Kraus channels \cite{kraus}, in step three. In the final step, the key characteristics of SPAD, such as efficiency, timing jitter, thermal dark counts, and afterpulsing, are analyzed and visualized as expectation values. The complete step-by-step methodology for the quantum simulation framework is illustrated in Figure \ref{fig:sim_steps}.

\subsection{Pauli Strings Mapping} \label{3.1}
While classical computers work on the binary numbers $\{0, 1\}$, quantum computing works on superpositions of these bits, within the $2^{n_q}$-dimensional Hilbert space defined by the computational basis $\{|0\rangle, |1\rangle\}^{\otimes n_q}$ \cite{nielsen, qcg}. In the context of quantum simulation, physical observables are mapped to Pauli strings, which are tensor products of Pauli matrices. The measurements are taken along the computational $Z$-basis and do not yield binary $\{0,1\}$ outputs, but they collapse the state into the eigenstates of the Pauli-Z operator, yielding the physical eigenvalues of $+1$ or $-1$. Hence, to simulate the SPAD detector on a quantum computer, the Hamiltonian of Eq.~(\ref{full}) is expressed as a Pauli string representation as,
\begin{equation} \label{Paulistring}
    H_{JCM} = \sum_{\text{P}_i \in \{I,X,Y,Z\}^{\otimes n}} c_i P_i
\end{equation}
where $c_i \in \mathbb{R}$ and is a scaler while $P_i$ represents Pauli strings. The Pauli strings are Hermitian and satisfy the orthogonality relation $\text{Tr}(P_i P_j) = 2^n \delta_{ij}$. Pauli string contains terms like X, X$\otimes$Z, Y$\otimes$Y$\otimes$Z.. etc. The length of one Pauli string depends upon the number of qubits required to simulate the problem. The set of Pauli string bases $\Sigma_1$ and $\Sigma_2$ for one and two-qubit systems is written as Eq.~(\ref{str1})  and (\ref{str2}).
\begin{equation} \label{str1}
\Sigma_1=\{I,X,Y,Z\}
\end{equation}
\begin{equation} \label{str2}
    \Sigma_2=\{II,IX,IY,IZ, XI, XX, XY, XZ,YI,YX,YY,YZ,ZI,ZX,ZY,ZZ\}
    \end{equation}
Where each term of the set $\Sigma_2$ is written as the tensor product of $\Sigma_1\otimes \Sigma_1$. Similarly, the Pauli string basis for an n-qubit system can be obtained as $\Sigma_n=\Sigma_1^{\otimes n}$. The coefficients $c_i$ of Eq.~(\ref{Paulistring}) are determined by the following expression \cite{fondana},
\begin{equation} \label{ci}
    c_i = \frac{1}{2^{n_q}} \text{Tr}(P_i\cdot H)
\end{equation}
where terms $P_i$ $\in$ $\Sigma$ sets described above in Eq. (\ref{str1}) and (\ref{str2}). To map the JC Hamiltonian into the Pauli strings, Qiskit's \texttt{SparsePauliOp} \cite{sp} is used, which internally decomposes the given matrix into Pauli strings, utilizing Eq.~(\ref{ci}). Furthermore, utilizing Eq.~(\ref{ci}) and the matrix of Eq.~(\ref{creation}), the Pauli string mapping for the Photon Hamiltonian described in Eq.~(\ref{jjs}) is expressed as,
\begin{equation}
    H_{\text{Photon}}=\hbar\Omega(\frac{3}{2}I_{p1}I_{p0}-\frac{1}{2} I_{p1}Z_{p0}-Z_{p1}I_{p0})
\end{equation}
where $p_1\ \&\ p_0$ represents qubits for the photon  Pauli matrix, with a subscript applied to the respective photon qubit. In writing the above Hamiltonian, the \textbf{Little-Endian notation} is taken into account, in which the quantum state layout is given as $\text{Photon}\otimes \text{SPAD}$, i.e., the SPAD qubits are placed from the top-most position in the quantum circuit \cite{qc} followed by the qubits assigned for Photon. 

The Pauli string mapping for the three-level SPAD Hamiltonian matrix described in Eq.~(\ref{ham}) is expressed as follows, 
\begin{equation}
\begin{split}
H_{\text{SPAD}} =
\left(\frac{E_g+E_e+E_t}{4}\right) I_{s1} I_{s0}
+ \left(\frac{E_g-E_e+E_t}{4}\right) I_{s1} Z_{s0} \\
+ \left(\frac{E_g+E_e-E_t}{4}\right) Z_{s1} I_{s0}
+ \left(\frac{E_g-E_e-E_t}{4}\right) Z_{s1} Z_{s0}
\end{split}
\end{equation}

Furthermore, the Pauli string mapping of the interaction Hamiltonian of Eq.~(\ref{16intmatrix}) is expressed as,
\begin{equation} \label{intpauli}
\begin{aligned}
H_{\text{int}} &= \hbar g \biggl[
\frac{1+\sqrt{3}}{4}
\big( I_{p3} X_{p2} I_{s1} X_{s0} + I_{p3} X_{p2} Z_{s1} X_{s0} + I_{p3} Y_{p2} I_{s1} Y_{s0} + I_{p3} Y_{p2} Z_{s1} Y_{s0} \big) \\
&\qquad + \frac{\sqrt{2}}{4}
\big( X_{p3} X_{p2} I_{s1} X_{s0} + X_{p3} X_{p2} Z_{s1} X_{s0}
- X_{p3} Y_{p2} I_{s1} Y_{s0} - X_{p3}Y_{p2} Z_{s1} Y_{s0} + Y_{p3} X_{p2} I_{s1} Y_{s0}\\ &\qquad + Y_{p3} X_{p2}Z_{s1} Y_{s0}
+ Y_{p3} Y_{p2} I_{s1} X_{s0} + Y_{p3} Y_{p2} Z_{s1} X_{s0} \big)  - \frac{\sqrt{3}-1}{4}
\big( Z_{p3} X_{p2} I_{s1} X_{s0} + Z_{p3} X_{p2} Z_{s1} X_{s0}\\
&\qquad
+ Z_{p3} Y_{p2} I_{s1} Y_{s0}+ Z_{p3} Y_{p2} Z_{s1} Y_{s0} \big)
\biggr]
\end{aligned}
\end{equation}

Each term of the Pauli string Hamiltonian of the interaction term contains four Pauli matrices in tensor product, such that the interaction Hamiltonian matrix is a $16\times 16$ matrix, but the Hamiltonian described for the Photon and the SPAD detector in Eqs.~(\ref{jjs}) and (\ref{ham}), respectively, would yield $4\times 4$ matrices. These Hamiltonians can not be summed with the interaction Hamiltonian directly. Hence, to account for the combined Hilbert space of Photon-SPAD, the individual four-dimensional Hilbert space of Photon and SPAD is promoted to a sixteen-dimensional Hilbert space, and it is described as follows,

\begin{equation} \label{16SPAD}
\begin{split}
H_{\text{SPAD}} =
\left(\frac{E_g+E_e+E_t}{4}\right) I_{p3} I_{p2} I_{s1} I_{s0}
+ \left(\frac{E_g-E_e+E_t}{4}\right) I_{p3} I_{p2} I_{s1} Z_{s0} \\
+ \left(\frac{E_g+E_e-E_t}{4}\right) I_{p3} I_{p2} Z_{s1} I_{s0}
+ \left(\frac{E_g-E_e-E_t}{4}\right) I_{p3} I_{p2} Z_{s1} Z_{s0}
\end{split}
\end{equation}
 \begin{equation} \label{16Photon}
    H_{\text{photon}}=\hbar\Omega \left(\frac{3}{2}I_{p3}I_{p2} I_{s1}I_{s0} - \frac{1}{2} I_{p3} Z_{p2}I_{s1}I_{s0} - Z_{p3}I_{p2} I_{s1}I_{s0}\right)
    \end{equation}

Eqs.~(\ref{intpauli}), (\ref{16SPAD}) and (\ref{16Photon}) combine to define the Hamiltonian $H_{JCM}$. The qubit layout for the combined quantum system to be fed into the quantum circuit is described in the table (\ref{tab:Qubit_layout}).
\begin{table}[H]
    \centering
    \renewcommand{\arraystretch}{1.3} 
    \setlength{\tabcolsep}{8pt}       
    \caption{Qubit layout for Quantum circuit}
    \label{tab:Qubit_layout}
    
    \begin{tabular}{ l l c } 
        \toprule
        \rowcolor{blue!10} 
        \textbf{Qubit} & \textbf{System} & \textbf{Symbol} \\
        \midrule
        $0^{th}$ & SPAD & $s_0$ \\
        $1^{st}$ & SPAD & $s_1$ \\
        $2^{nd}$ & Photon & $p_2$ \\
        $3^{rd}$ & Photon & $p_3$ \\ 
        \bottomrule
    \end{tabular}
\end{table}
\subsection{Quantum Circuits and states}
The Quantum simulation is initialized with the quantum circuit \cite{qc}. A classical circuit has wires and charges to carry signals, but Quantum circuits don't have physical wires. Instead, in their pictorial view, there are many horizontal lines that represent the qubit in any Quantum circuit. To manipulate the state of a qubit in the quantum circuit, Quantum gates are used. Quantum gates are unitary matrices that act on the quantum state and transform it \cite{nielsen, qcg}. The Quantum circuits used in this simulation are described in this section.
\subsubsection{Initial Quantum states}
\begin{figure}[H]
    \centering
    \begin{quantikz}[row sep=0.4cm, column sep=0.6cm]
        \lstick{$SPAD\_{0}$} & \qw & \qw \\
        \lstick{$SPAD\_{1}$} & \qw & \qw \\
        \lstick{$ph\_{1}$}   & \gate[style={fill=blue!20}]{X} & \qw \\
        \lstick{$ph\_{2}$}   & \gate[style={fill=blue!20}]{X} & \qw
    \end{quantikz}
    \caption{Quantum circuit for simulating Photon-SPAD interaction. The initial state is $\ket{1100}$.}
    \label{fig:spad_3level_1}
\end{figure}
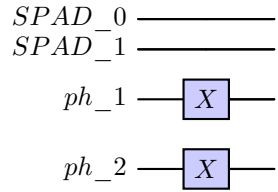

A quantum circuit is always initialized in the vacuum state \cite{jjs}, i.e., the initial state is $\ket{000...000}$, without any operations. To initialize the system with a definite number of photons and SPADs in a specific state, the discussion of standard binary encoding in sections \ref{SPADSBE} and \ref{PFQ} is used.
If there are n photons in the system, and the SPAD is in the ground state ($\ket{g}$), then the quantum state of this system is written as $\ket{n}\otimes\ket{g}$. For n = 3, the quantum state in the binary encoding is written as $\ket{11}\otimes\ket{00}$. To represent this quantum state in a quantum simulation, the Pauli X gate is applied to both of the photon qubits. Thus, the quantum circuit for this initial state is constructed as the figure \ref{fig:spad_3level_1}.

The specific quantum circuit of figure \ref{fig:spad_3level_1} is used to simulate SPAD efficiency.
Furthermore, if the SPAD detector is in the excited state and there are no photons in the system, the Pauli X gate is applied to the first SPAD qubit. The quantum circuit for this scenario is shown in figure \ref{fig:spad_side_by_side}(a). Furthermore, when the SPAD detector is in the trap state with no photon in the system, then the quantum state of the system is $\ket{0010}$. The quantum circuit for this state is shown in figure \ref{fig:spad_side_by_side}(b). This quantum circuit is used in the afterpulsing analysis. A quantum circuit is shown in the figure \ref{fig:spad_side_by_side}(c), where there exist no gates to alter the initial state. The quantum state of this system is kept at the default vacuum state, i.e., $\ket{0000}$. This quantum circuit represents the initial state to simulate the thermal dark counts of the SPAD detector.
\begin{figure}[H]
    \centering

    \begin{minipage}{0.30\textwidth}
        \centering
        \begin{quantikz}[row sep=0.4cm, column sep=0.6cm]
            \lstick{$SPAD\_{0}$} & \gate[style={fill=blue!20}]{X} & \qw \\
            \lstick{$SPAD\_{1}$} & \qw & \qw \\
            \lstick{$ph\_{1}$}   & \qw & \qw \\
            \lstick{$ph\_{2}$}   & \qw & \qw
        \end{quantikz}

        \caption*{(a) When the initial quantum state is $\ket{0001}$.}
        \label{fig:spad_3level_2a}
    \end{minipage}
    \hfill
    \begin{minipage}{0.30\textwidth}
        \centering
        \begin{quantikz}[row sep=0.4cm, column sep=0.5cm]
            \lstick{$SPAD\_{0}$} & \qw & \qw \\
            \lstick{$SPAD\_{1}$} & \gate[style={fill=blue!20}]{X} & \qw \\
            \lstick{$ph\_{1}$}   & \qw & \qw \\
            \lstick{$ph\_{2}$}   & \qw & \qw
        \end{quantikz}

        \caption*{(b) When the initial quantum state is $\ket{0010}$.}
        \label{fig:spad_3level_2b}
    \end{minipage}
    \hfill
    \begin{minipage}{0.30\textwidth}
        \centering
        \begin{quantikz}[row sep=0.6cm, column sep=0.5cm]
            \lstick{$SPAD\_{0}$} & \qw & \qw \\
            \lstick{$SPAD\_{1}$} & \qw & \qw \\
            \lstick{$ph\_{1}$}   & \qw & \qw \\
            \lstick{$ph\_{2}$}   & \qw & \qw
        \end{quantikz}

        \caption*{(c) When the initial quantum state is $\ket{0000}$.}
        \label{fig:spad_3level_2c}
    \end{minipage}

    \caption{Quantum circuits representing different quantum circuit configurations for different initial states.}
    \label{fig:spad_side_by_side}
\end{figure}
Once the quantum circuit is initialized, the observable states are selected, depending on the SPAD detector property being simulated. The Eq~(\ref{jcref}) is used for this purpose. In the appendix (\ref{appB}), the SPAD detector observable states, with their physical meaning are defined. 
\subsection{Time Evolution and Trotterization}

\begin{figure}[H]
     \centering
     \includegraphics[width=0.8\linewidth]{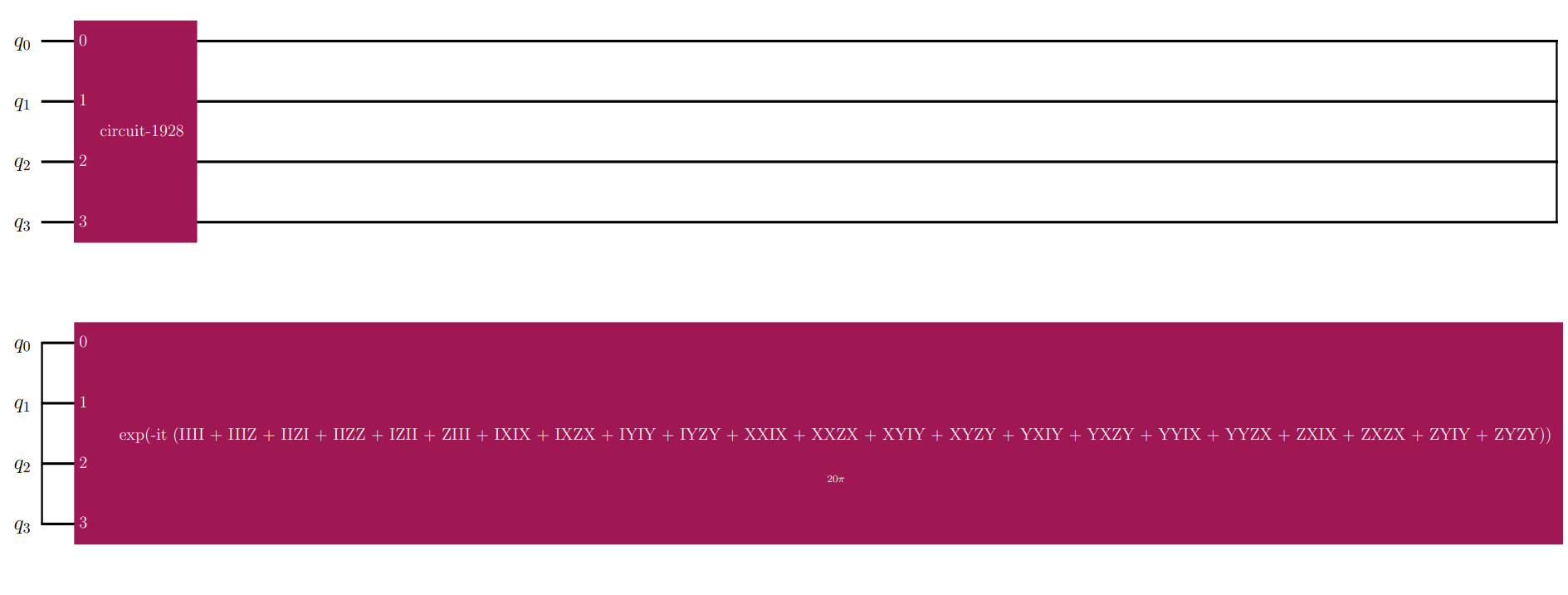}
     \caption{Quantum circuit for one Trotter step.}
     \label{fig:trottercirc}
 \end{figure}

To simulate the key characteristics of the SPAD system, the initial quantum state is evolved over time. The first-order Lie-Trotter-Suzuki approximation, known as Trotterization \cite{suzuki1991, trotterQ}, is used to address non-commutation and simulate the time evolution\footnote{See appendix \ref{appC} for a detailed discussion.}. This method resolves this limitation by slicing the total simulation time $t$ into $n$ discrete, small time intervals $\Delta t = t/n$. For infinitesimally small time steps, the commutator error becomes negligible, allowing the unitary operator to be approximated as a sequential product of sub-evolutions of $\Delta t$ time. Hence, the time-evolution operator is written as in Eq.~(\ref{trotter}).
\begin{equation} \label{trotter}
    U(t) \approx \left( \prod_{i} \exp\left(-\frac{i c_i P_i \Delta t}{\hbar}\right) \right)^n
\end{equation}
Hence the evolved state $\psi(t)$ can be approximated as,
\begin{equation} \label{psifinal}
    \vert \psi(t) \rangle \approx \left( \prod_{i} \exp\left(-\frac{i c_i P_i \Delta t}{\hbar}\right) \right)^n \vert\psi(0) \rangle
\end{equation}
 As $\Delta t \to 0$, this iterative sequence converges to the true continuous dynamics of the system \cite{suzuki1991}. A Quantum computer uses Rotation Gates to implement Trotterization. In the Qiskit implementation, each exponential term in Eq.~(\ref{trotter}) is compiled into parameterized single-qubit rotations ($R_x, R_y, R_z$) and two-qubit entangle gates (CNOT). The Hamiltonian term ($H_{JCM}$) determines the Axis of Rotation (X, Y, or Z). The time step ($\Delta t$) determines the Angle of Rotation ($\theta$). Figure \ref{fig:trottercirc} shows one full trotter step. The initial circuit block labeled "circuit 1928"\footnote{"circuit 1928" is nothing to do with the quantum simulation; it is just the syntax used by Qiskit to mark the number of quantum circuits constructed by the user in a notebook.} represents the initial state of the quantum system, followed by the Trotterization block. Furthermore, in this simulation, the number of Trotter steps $n$ is set to 200, which means the quantum circuit shown in Figure \ref{fig:trottercirc} is repeated 200 times. The decomposed circuit for better visualization is presented in the appendix (\ref{appD}).
 
 To perform Trotterization in Qiskit, an instance \texttt{TrotterQRTE} is created, which is in the class \texttt{TimeEvolutionProblem} \cite{trotterQ}, imported from \texttt{qiskit\_algorithms} \cite{Qalgo}.

\subsection{Open Quantum system dynamics of the SPAD}
\label{sec:open_quantum}

The Jaynes-Cummings Hamiltonian describes a perfectly isolated, closed system in which only the photon-mediated excitation between $\ket{g}\ \&\ \ket{e}$ is modeled.  However, a real SPAD detector is never isolated. It is continuously coupled to the external environment, such as external temperature, radiation fluence, and crystal lattice defects. These couplings reduce efficiency and increase timing jitter, dark counts, and afterpulsing. To accurately model SPAD performance parameters, the open quantum systems \cite{OQS, Lidar2019} framework is used, which captures the combined effects of coherent dynamics and irreversible environmental noise. The Lindblad master equation \cite{GKLS, GKLS1, GKLS2, manzano} and Qiskit's noise channels, such as thermal relaxation \cite{qn} and Kraus channel \cite{kraus}, are used in this quantum simulation to simulate SPAD detector. 
 

\subsubsection{Jump Operators for the Three-Level SPAD}
\label{sec:jump_ops}
The time evolution of the Photon-SPAD density matrix $\rho(t)$ in the presence of environmental noise is governed by the Lindblad master equation\footnote{Also known as Gorini--Kossakowski--Sudarshan--Lindblad (GKSL) master equation, independently derived by Gorini, Kossakowski, and Sudarshan~\cite{GKLS1}
and by Lindblad~\cite{GKLS2}, both in 1976}. This equation is expressed in Eq.~(\ref{GKLS}).
 
\begin{equation} \label{GKLS}
    \frac{d\rho}{dt} = -i[H, \rho] + \sum_k \gamma_k \left( L_k \rho L_k^\dagger - \frac{1}{2}\{L_k^\dagger L_k, \rho\} \right)
\end{equation}
The operator $L_k$ is a jump operator. It encodes the environment-induced transitions, such as $\ket{g} \to \ket{e}$ for a dark count, and $\ket{t} \to \ket{e}$ for afterpulsing. $\gamma_k \geq 0$ is the rate at which that process occurs
~\cite{OQS, Lidar2019}.
The dimension of each operator depends on the chosen truncation. If the photon field is truncated at $N_\text{max}$ photons, the annihilation operator $a$ is an $(N_\text{max}+1)\times(N_\text{max}+1)$ matrix. For the three-level SPAD, it is $4\times 4$, because the three physical states $\{\ket{g}, \ket{e}, \ket{t}\}$ are embedded in the Hilbert space of two qubits (dimension 4), with the fourth computational basis state unused. The full jump operators therefore live in a Hilbert space of dimension $4(N_\text{max}+1)$, and the density matrix $\rho$ has dimension $[4(N_\text{max}+1)]^2$. Hence, the full Hilbert space of the SPAD-photon system is a tensor product $\mathcal{H}_\text{ph} \otimes \mathcal{H}_\text{SPAD}$.
Each jump operator $L_k$ is therefore constructed as a tensor product of a photon-space operator and a SPAD-space operator as,
\begin{equation} \label{L_k}
    L_k = \hat{O}_\text{ph}^{(k)} \otimes \hat{O}_\text{SPAD}^{(k)}
\end{equation}
For the two-level SPAD ($\ket{g}$, $\ket{e}$), the SPAD operators are the standard Pauli raising/lowering operators $\sigma_\pm$.
For the three-level SPAD ($\ket{g}$, $\ket{e}$, $\ket{t}$), they are explicit transition operators defined as
\begin{equation} \label{jumpop}
    \sigma_{ij} \equiv \ket{i}\bra{j}, \qquad i,j \in \{g, e, t\}
\end{equation}
where the subscript gives the direction of the transition, for example, $\sigma_{eg}$ takes the system from ground state ($\ket{g}$) to excited state ($\ket{e}$). Furthermore, the states $\ket{g},\ \ket{t}\ \&\ \ket{e}$ are written in the two-qubit computational basis states described in Eq.~(\ref{matrixf}).

 \subsubsection{Gate-Based Noise Model in Qiskit}
\label{sec:gate_noise}
 
The Lindblad master equation of Eq.~(\ref{GKLS}) provides a
continuous-time description of SPAD noise. However, the evolution in a quantum circuit is discrete, and noise is included as an error
channel, applied as quantum gates. In this section, the simulation of thermal dark counts and afterpulsing in the discrete, gate-based framework using Qiskit~\cite{qiskit} is described. The thermal dark counts are simulated using \texttt{thermal\_relaxation\_error} \cite{qn} while afterpulsing is simulated using
\texttt{Kraus} channel \cite{kraus}.
 To simulate the thermal dark counts, the quantum circuit is initialized as figure \ref{fig:spad_side_by_side}(c), and the probability of the spontaneous transitions from its ground state $\ket{g}$ to its excited state $\ket{e}$ due to the thermal energy is measured
In this process. The rate at which dark counts are thermally generated is given by the
Arrhenius equation~\cite{IEEE, Sze2006}.
\begin{equation} 
    R(T) = A_\text{eff}
           \exp\!\left( -\frac{\Delta E}{k_B T} \right)
    \label{eq:arrhenius}
\end{equation}
where $\Delta E$ is the activation energy for carrier generation,
$k_B$ is Boltzmann's constant, $T$ is the temperature in kelvin, and $A_\text{eff}$ is the attempt frequency.
The \texttt{thermal\_relaxation\_error} has four parameters, such as relaxation time ($t_1$), dephasing time ($t_2$), gate time, and excited state population ($p_{th}$). $t_1$ is the average time for the excited state $\ket{e}$ to decay back to $\ket{g}$. The Arrhenius rate $R(T)$ is mapped to the qubit's relaxation parameters as described in Eq.~(\ref{eq:T1T2}).
\begin{equation}
    t_1 = \frac{1}{R(T)}, \qquad t_2 = 2\,t_1
    \label{eq:T1T2}
\end{equation}
The thermal excited-state population ($p_\text{th}$) is the probability of finding the qubit in $\ket{e}$ at the given temperature $T$ and is given by the Fermi-Dirac statistics as,
\begin{equation}
    p_\text{th}(T)
    = \frac{1}{1 + \exp\!\left(\dfrac{\Delta E}{k_B T}\right)}
    \label{eq:pth}
\end{equation}

Furthermore, afterpulsing in the SPAD is simulated using the quantum circuit shown in
Figure~\ref{fig:spad_side_by_side}(b). Physically, a carrier trapped in state $\ket{t}$
after an avalanche event can later transition to either $\ket{e}$ or $\ket{g}$, and this
process is described using two independent sub-events~\cite{dsouza}. The first is detrapping, the carrier escapes the defect with a time-dependent probability
$P_\text{release}(t)$. The second is avalanche triggering, once the charge carrier is de-trapped it has some probability
$P_\text{ap}$ of actually producing a detectable avalanche rather than relaxed to the ground state $\ket{g}$. Hence, the overall afterpulse probability is given by,
\begin{equation}
    P_\text{avalanche}(t) = P_\text{release}(t) \times P_\text{ap}
    \label{eq:branch}
\end{equation}
Afterpulsing is simulated with a Kraus channel acting on the SPAD subspace ($q_0,q_1$), with the trap state branching into three
possible outcomes each timestep as,
\begin{align}
    K_{\mathrm{av}}   &= \sqrt{P_{\mathrm{release}}.\, P_{\mathrm{ap}}}\;
                         \ket{e}\bra{t}, &&\ket{t}\to\ket{e}\ \text{(avalanche triggered)},
    \label{eq:kav}\\
    K_{\mathrm{safe}} &= \sqrt{P_{\mathrm{release}}\,(1-P_{\mathrm{ap}})}\;
                         \ket{g}\bra{t}, &&\ket{t}\to\ket{g}\ \text{(safe recombination)},
    \label{eq:ksafe}\\
    K_{\mathrm{stay}} &= \sqrt{1-P_{\mathrm{release}}}\;
                         \ket{t}\bra{t} + \sum_{i\neq t}\ket{i}\bra{i},
                         &&\ket{t}\to\ket{t}\ \text{(remains trapped)}.
    \label{eq:kstay}
\end{align}
These satisfy the completeness condition
$K_{\mathrm{av}}^\dagger K_{\mathrm{av}} + K_{\mathrm{safe}}^\dagger K_{\mathrm{safe}}
+ K_{\mathrm{stay}}^\dagger K_{\mathrm{stay}} = \mathbf{I}$\footnote{See appendix \ref{app:trace_preservation} for its derivation.}, ensuring trace
preservation~\cite{kraus}.

Both terms in Eq.~(\ref{eq:branch}) evolve with radiation damage, but for different
physical reasons. Experimentally, $P_\text{ap}$ increases almost linearly with fluence
$\phi$~\cite{anisimova, IEEE}, since more radiation-induced traps mean a larger fraction
of released carriers succeed in triggering an avalanche. Writing $P_{\text{ap},0}$ as the
afterpulse probability measured at a reference fluence $\phi_0$, the fluence-scaled value
follows the first-order approximation~\cite{anisimova} as,
\begin{equation}
    P_\text{ap}(\phi) = P_{\text{ap},0} \times \frac{\phi}{\phi_0}.
    \label{eq:pap}
\end{equation}
The high-energy space radiation produces deep-level defect energy levels. Hence the trap life time of charge carriers are different. A carrier's escape probability from the trap level $\ket{t}$ depends on the time it has been trapped and this aggregate behavior is modeled with a Generalized Fractional
Poisson (GFP) process~\cite{afterpulsing25}, giving the power-law survival
function~\cite{wei2024, wang2025}
\begin{equation}
    S_{\mathrm{PL}}(t) = \left(\frac{\tau_0}{\tau_0 + t}\right)^{\!\alpha}, \qquad \alpha > 0,
    \label{eq:survival_pl}
\end{equation}
where $\tau_0$ is the characteristic trap lifetime and $\alpha$, which is the power law coefficient, controls the tail of the
distribution. Discretizing this over a simulation timestep $\Delta t$, the
conditional release probability at elapsed time $t_n = n\Delta t$ is described as,
\begin{equation}
    P_{\mathrm{release}}(t_n) = 1 - \frac{S_{\mathrm{PL}}(t_n + \Delta t)}{S_{\mathrm{PL}}(t_n)}
    = 1 - \left( \frac{\tau_0 + t_n}{\tau_0 + t_n + \Delta t} \right)^{\!\alpha},
    \label{eq:prelease_pl}
\end{equation}
which is used directly into the Kraus channels. Furthermore, The Lindblad rates and various simulation parameters used in this quantum simulation can be found in appendix~(\ref{appF}).

\section{Results and Discussion}
In this section, the quantum simulation results for key characteristics of the SPAD detector, such as efficiency, timing jitter, thermal dark counts and afterpulsing are described in detail, using Quantum simulations
\subsection{Efficiency of the SPAD detector}
The Efficiency of the SPAD detector is a key metric for single-photon detection in QKD, LiDAR, and various imaging tasks. It is defined as the probability that a SPAD absorbs a photon and produces an avalanche. In this simulation, the Efficiency is measured by calculating the time-dependent probability of the observable $\ket{n-1, e}\bra{n-1,e}$, which implies a photon-mediated transition that couples the two quantum states $\ket{n,g}$ and $\ket{n-1, e}$, and produces characteristic Rabi oscillations. This phenomenon corresponds to the specific quantum state in which a photon is absorbed and the photon Fock state becomes $\ket{n-1}$ from the initial state $\ket{n}$, and the SPAD makes a transition to the excited state $\ket{e}$ from the ground state $\ket{g}$ (\ref{fig:three_level_spad}). The observable states that are used in this simulation corresponding to their initial state are described in Eq.~(\ref{eq:obsquan}),
\begin{equation} \label{eq:obsquan}
    \begin{aligned}
        & \textbf{Resonant Transitions } (|n, g\rangle \leftrightarrow |n-1, e\rangle): \\
        & |1, g\rangle = |0100\rangle \longleftrightarrow |0, e\rangle = |0001\rangle \\
        & |2, g\rangle = |1000\rangle \longleftrightarrow |1, e\rangle = |0101\rangle \\
        & |3, g\rangle = |1100\rangle \longleftrightarrow |2, e\rangle = |1001\rangle
    \end{aligned}
\end{equation}
\subsubsection{Ideal (closed system) Efficiency}
\label{sec:ideal_efficiency}
\begin{figure}[H]
   
        \centering
        \includegraphics[width=0.7\linewidth]{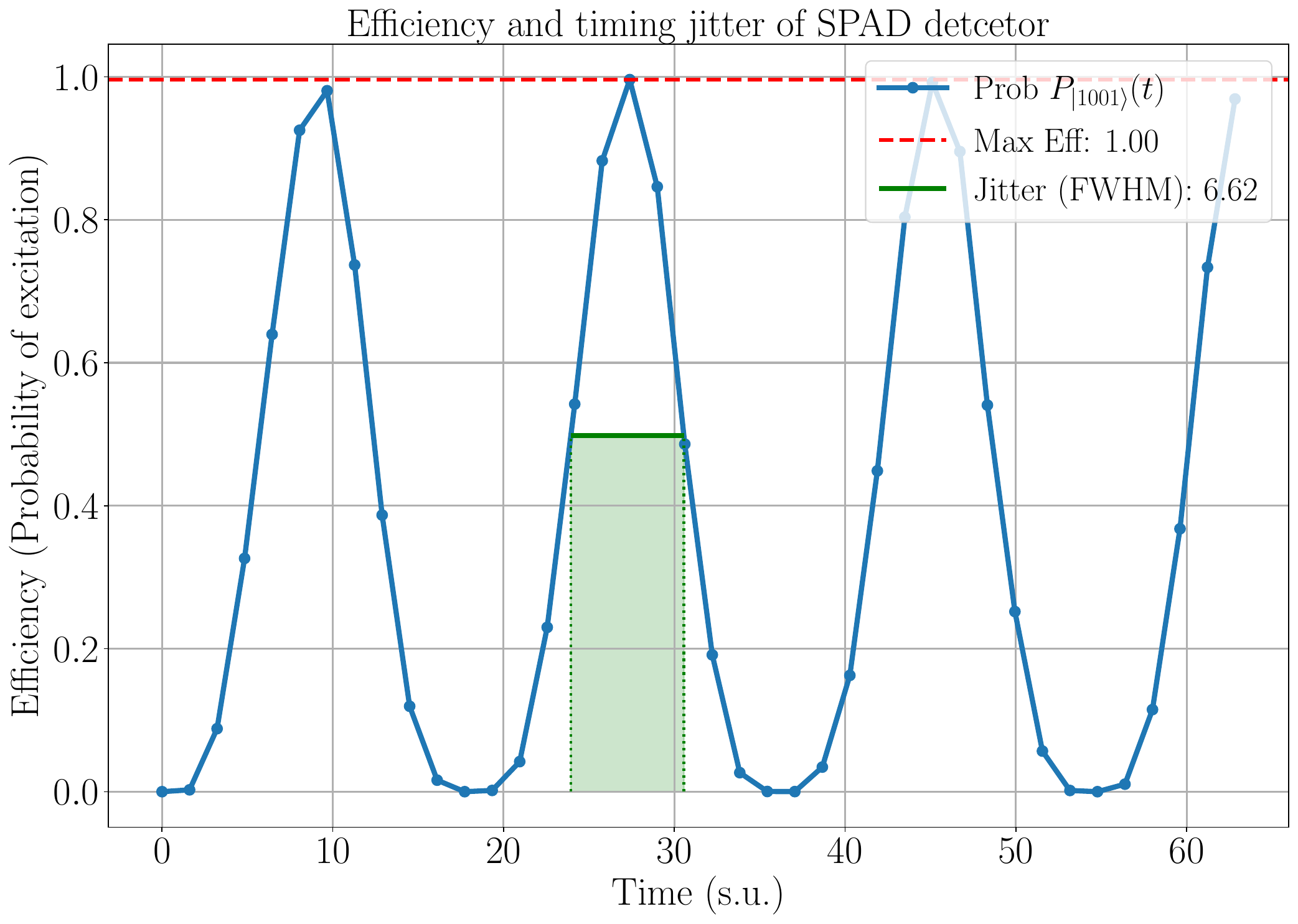}
        \caption{The figure shows coherent Rabi oscillations and efficiency and timing jitter of the SPAD detector is simulated in Qiskit's Aer simulator. All time units are in simulation units, described in the appendix.}
        \label{fig:circuit_jitter}
\end{figure}

The main goal of the ideal simulation is to establish a baseline for SPAD detection efficiency under the Jaynes-Cummings interaction alone, without introducing environmental effects.
This allows us to identify the intrinsic quantum limit of detection
efficiency.
Figure~(\ref{fig:circuit_jitter}) shows the detection probability
$P_{|1001\rangle}(t)$ as a function of time, simulated for the ideal,
closed three-level SPAD system using Qiskit's Trotterized time evolution
(\texttt{TrotterQRTE}, 200 Trotter steps).
The system is initialized in the state with three photons and the SPAD in the ground state $\ket{g}$, i.e., $\ket{1100}$. The observable is the projector onto the target state $\ket{1001}$, measuring the probability that the photon has been absorbed and the detector has transitioned to its excited state $\ket{e}$, i.e. $|\bra{1001}U|\ket{1100}|^2$, where U is the Trotterized time evolution operator expressed in Eq.~(\ref{trotter}). The dashed red horizontal line marks the maximum detection probability
achieved. Furthermore, the green shaded region of figure (\ref{fig:circuit_jitter}) shows the full-width-at-half-maximum (FWHM) of the probability curve, which is used to estimate the timing jitter.

The probability curve exhibits periodic oscillations between zero and a maximum. This behavior is the well-known Rabi oscillation of the
Jaynes-Cummings model~\cite{jcmodel, swine}. The Rabi frequency is given by,
$\Omega_R = g\sqrt{n+1}$, where $g = 0.1$  is the coupling constant measured in simulation units, and
$n = 3$ is the photon number.
At $t = 0$, the system starts with three photons in the field and the atom in the ground state, so the detection probability is zero. As time passes, one photon is annihilated, and the SPAD transitions to the excited state, $\ket{e}$. It reaches its peak when the photon is completely annihilated, then the SPAD de-excites.
The oscillation repeats at the Rabi period $T_R = 2\pi/g$. This behavior is coherent, meaning there is no irreversible process and no energy is lost. The photon simply oscillates back and forth between the field and the detector indefinitely. This represents the quantum superposition of
$\ket{3,g}$ and $\ket{2,e}$ states, where the system periodically oscillates each configuration~\cite{Gerry2005}.
 
The detection efficiency of the SPAD detector is given by the amplitude value of the probability curve shown in the figure (\ref{fig:circuit_jitter}), is turned out to be approximately 1, which is the highest probability
achievable through the Jaynes-Cummings interaction alone. The full width at half maximum (FWHM) of the probability curve corresponds to the timing jitter of the SPAD detector. From the simulation, the FWHM is found to be,
\begin{equation}
    \Delta t_\text{FWHM} = 6.62 \text{ simulation time units.}
    \label{eq:fwhm_sim}
\end{equation}

This is a dimensionless number in the natural units of the Jaynes-Cummings model. To convert it to a physical time in seconds, a physical value needs to be assigned to the coupling constant $g$ used in the
simulation. In this simulation, the coupling constant, $g = 0.1$ is in simulation units, meaning one simulation
time unit equals $1/g_\text{phys}$ in seconds. Hence, expressing the simulation time units in seconds, the timing jitter obtained is 887 ps\footnote{See Appendix \ref{appE} for derivation of simulation unit.}. The simulated jitter is larger than the values measured in
real SPADs, and this is expected as, in a real SPAD, an active quenching circuit collapses the avalanche
very quickly. In this ideal Jaynes-Cummings simulation, there is no quenching
mechanism at all. The photon simply oscillates between the field and the detector at the Rabi frequency.

\subsubsection{Efficiency in open system}
\label{sec:noisy_efficiency}
\begin{figure}[H]
     \centering
     \includegraphics[width=0.6\linewidth]{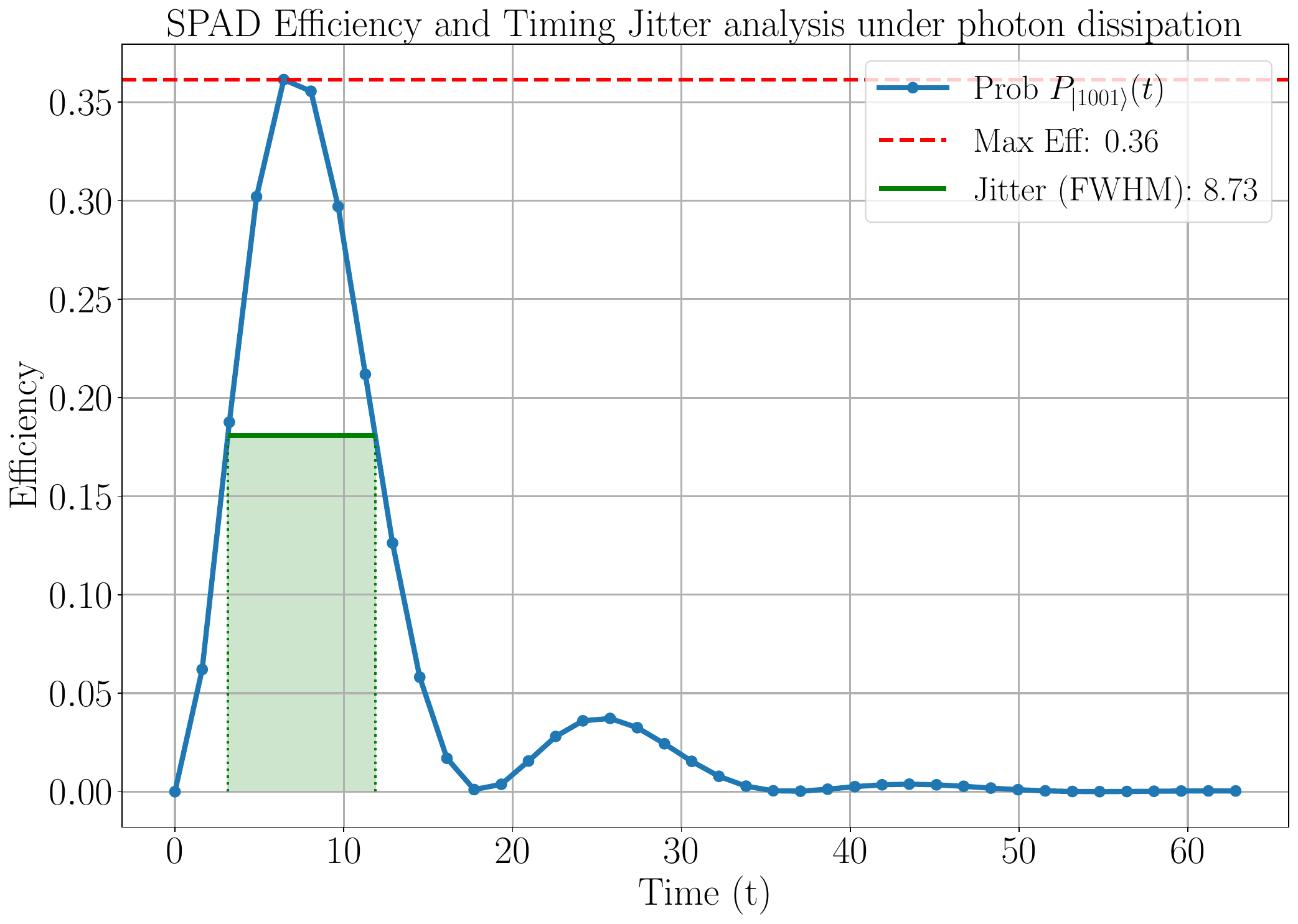}
     \caption{The detection probability computed using the Lindblad master equation, with photon loss (dissipation) channel.}
     \label{fig:noisy_efficiency}
 \end{figure}

The purpose of this simulation is to quantify how the photon dissipation degrades the
detection efficiency of the SPAD.
Photon dissipation models the absorption or scattering of the signal
photon by the medium before it reaches the active junction of the detector,
as well as cavity leakage in any resonator-coupled architecture \cite{jcmodel}.
In the presence of external environmental dissipation, a photon can be absorbed, scattered, or leak into the environment before interacting with the detector. It directly reduces the probability of detection. Figure~(\ref{fig:noisy_efficiency}) shows the detection probability
$P_{|1001\rangle}(t)$ described in the open noisy environment, computed using the Lindblad master equation,
with a single photon dissipation channel included via the jump
operator $L_\kappa = \sqrt{\kappa}\,(a \otimes I_\text{SPAD})$,
where $\kappa$ is the dissipation rate (or the coupling strength between photon and environment) which is $0.05$ in simulation units. The coupling strength between the noisy open environment is assumed to be half of the coupling strength $g$ of the ideal case.
The system is initialized in the state $\ket{1100}$, three photons
in the field ($N_\text{max} = 3$ truncation) and the detector in its
ground state.
The noisy probability curve shows three distinct physical features compared to the ideal case. Firstly, the peak efficiency is reduced by approximately two-thirds of the ideal efficiency (0.36). In the presence of dissipation, the photon may be absorbed or scattered at a rate $\kappa$ before it can produce the avalanche. The photon population in the field decays exponentially as
$\langle n(t)\rangle \propto e^{-\kappa t}$, hence, the probability of the SPAD detector being found in the excited state is decreased. Secondly, the system shows damped Rabi oscillations. The successive peaks become progressively smaller rather than remaining at constant height as in the ideal case. Physically, each time the photon returns to the field during a Rabi cycle, it has a nonzero probability of being lost to the environment before the
next absorption attempt. After many cycles, the photon is dissipated, and the system relaxes to the ground state with no detection event registered. Thirdly, the timing jitter also degrades as shown in the figure (\ref{fig:noisy_efficiency}), as the Rabi oscillations become fatter, the timing jitter gets poorer, 8.73 simulation units. Table~\ref{tab:efficiency_summary} summarises the key quantitative
results extracted from both simulations.
 
\begin{table}[H]
    \centering
    \renewcommand{\arraystretch}{1.1}
    \setlength{\tabcolsep}{8pt}
    \caption{Comparison of key efficiency metrics between the ideal
             (closed system, Trotterization) and noisy (open system,
             Lindblad) SPAD simulations.
             }
    \label{tab:efficiency_summary}
    \begin{tabular}{l c c}
        \toprule
        \rowcolor{blue!10}
        \textbf{Metric} &
        \textbf{Ideal simulation} &
        \textbf{Noisy simulation} \\
        \midrule

        Efficiency &
        1 &
       0.34 \\
        Timing jitter (FWHM) &
        877 ps &
        1169 ps \\
        Long-time behavior &
        Periodic Rabi oscillation &
        Decay to zero (photon dissipation) \\
        \bottomrule
    \end{tabular}
\end{table}

 \subsection{Thermal Dark Counts}
\label{sec:results_dark_counts}

\begin{figure}[H]
    \centering
    
   
        \includegraphics[width=1\linewidth]{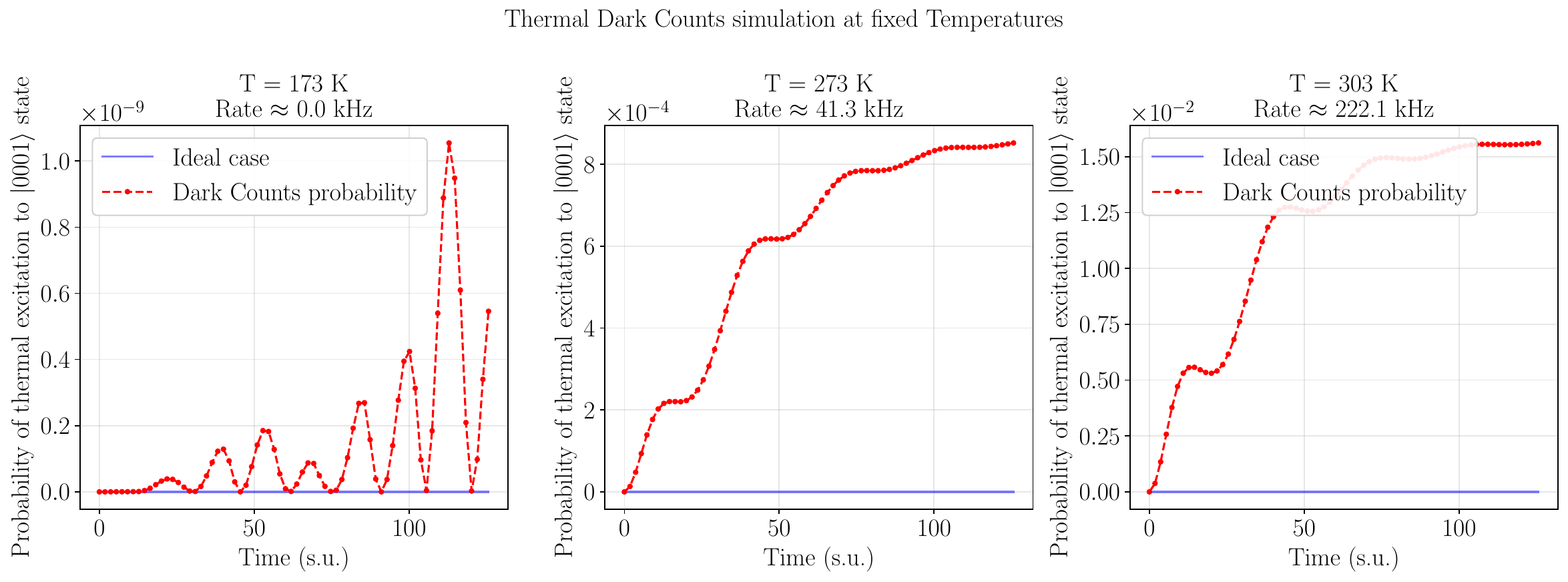}
        \caption{Thermal dark count probability with time, simulated by the Lindblad master equation.}
        \label{fig:Lindblad_DCR}
   
\end{figure}

\begin{figure}[H]
    \centering
    \includegraphics[width=1\linewidth]{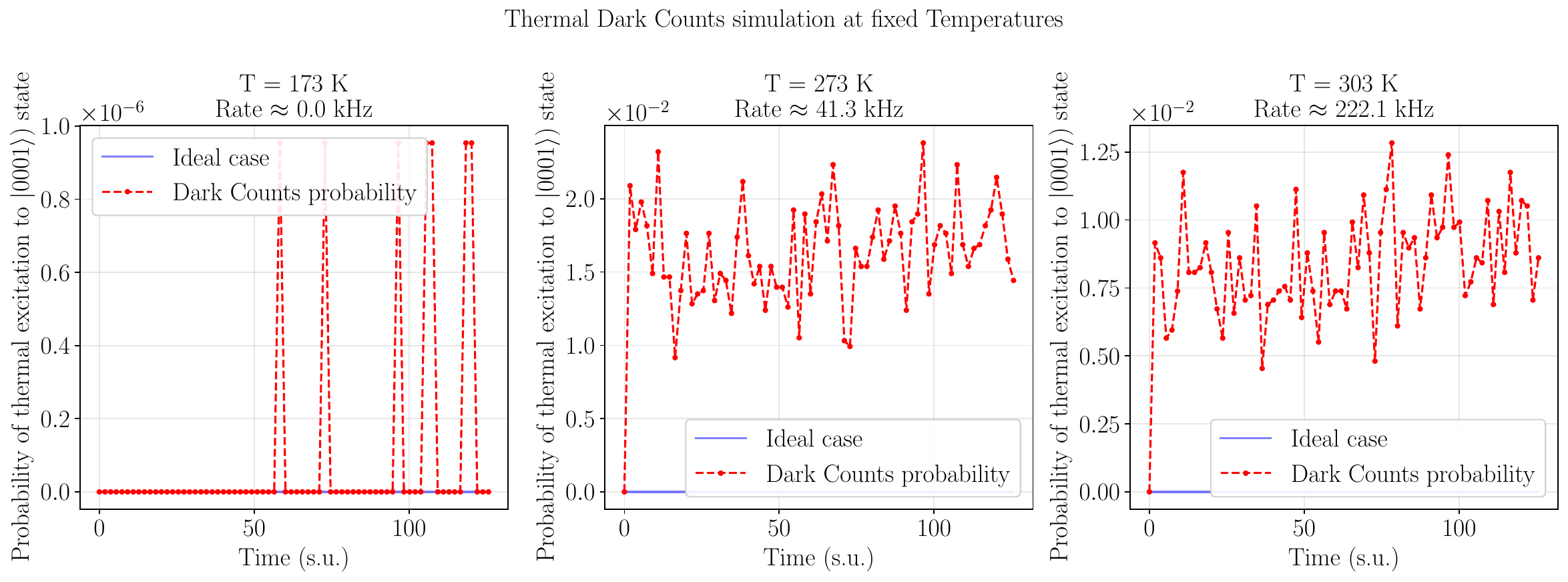}
    \caption{Gate-based Qiskit simulation of thermal dark counts in a 3-level SPAD model.}
    \label{fig:Qiskit_DCR}
\end{figure}

A dark count is a false detection event when the SPAD detector is excited to the $\ket{e}$ state in the absence of a photon. The thermal energy absorbed from the surroundings excites the carrier across the energy barrier i.e from the ground state $\ket{g}$, which is written as $\ket{0000}$ in computational basis, to the excited state/avalanche state $\ket{e}$, i.e. state $\ket{0001}$ in computational basis, without any photon presence. In this model, the thermal dark count is captured by two approaches. First, by the Lindblad master equation, in which the jump operator $L_\text{th} = \sqrt{\gamma_\text{th}}\,(I_\text{ph} \otimes
\sigma_{eg})$, is used to simulate the DCR, where the rate $\gamma_\text{th}$ is determined by the
Arrhenius equation described in Eq.(\ref{eq:arrhenius})~\cite{IEEE, Sze2006}. Second, IBM Qiskit's gate-based noise model, \texttt{thermal\_relaxation\_error} \cite{qn} is used, in which the system evolution is governed by the 4-qubit Hamiltonian solved via Trotterization (200 steps). Figures~(\ref{fig:Lindblad_DCR}) and~(\ref{fig:Qiskit_DCR}) show the
dark count probability $P(\ket{0001})(t)$ from the Lindblad and Qiskit simulations, respectively.

At a low temperature of 173\, K, both figures~(\ref{fig:Lindblad_DCR}) and~(\ref{fig:Qiskit_DCR}) show almost zero probability of the thermal dark counts, with a DCR of $\approx$ 2 counts per second. The SPAD detector shows almost ideal behavior at this temperature, confirming that cooling below $\sim$200\, K effectively eliminates thermal dark counts. Furthermore, at higher temperatures of 273\, K, and 303 K, a small non-zero probability appears at $t=0$ in both figures (\ref{fig:Lindblad_DCR}) and~(\ref{fig:Qiskit_DCR}). At higher temperatures, charge carriers gain enough energy to be liberated from the covalent bonds in the SPAD detector and reach to the excited state, and the detector has a measurable probability of being in the $\ket{e}$ state even in the absence of a photon in the system. At these higher temperatures, the thermal noise dominates, and the dark count rate is in the range of $\approx$ 41 to 222 kHz. 

It is observed from figures (\ref{fig:Lindblad_DCR}) and (\ref{fig:Qiskit_DCR} that the effect of the temperature on dark counts is exponential, not linear. Cooling a Si-SPAD from 303\, K to 173\, K, which is only 130\, K, reduces the DCR by 5/6 factor. Cooling further to 173\, K, as naturally occurs in low Earth
orbit~\cite{anisimova}, reduces it to very low values, making them well-suited for satellite QKD.
Both Lindblad and Qiskit simulations of thermal dark counts in figures (\ref{fig:Lindblad_DCR}) and~(\ref{fig:Qiskit_DCR}) show saturation in thermal dark count probability, which is the same behavior as observed in previous experimental studies \cite{anisimova}. The differences between the two approaches arise from the discretizations, the Lindblad solver integrates continuously, while the Qiskit model applies noise in discrete gate steps. These differences are negligible for $t_\text{gate} \ll t_1$, which
holds throughout the temperature range.
This agreement validates both implementations and confirms that the results are genuine physical prediction. 


\subsection{Afterpulsing}
\label{sec:results_afterpulsing}

\begin{figure}[H]
    \centering
    \begin{tikzpicture}[
        >=stealth,              
        node distance=3cm,      
        state/.style={          
            circle, 
            draw=black, 
            thick, 
            minimum size=1.0cm,
            font=\large\bfseries
        },
        every edge/.style={     
            draw, 
            thick, 
            ->
        }
    ]

    \node[state] (e) at (0,0) {$|e\rangle$};

    \node[state] (g) at (6,0) {$|g\rangle$};

    \node[state] (t) at (3,-2.5) {$|t\rangle$};


    \path (e) edge node[above, font=\small] {Relaxation} (g);

    \path (e) edge [bend right=20] node[left, xshift=1.6cm, font=\small, align=right] {Trapping} (t);

    \path (t) edge [bend right=20] node[right, xshift=0.26cm, font=\small, align=left] {Upon Releasing\\(Normal Detrapping)} (g);

    \path (t) edge [bend left=135, dashed, color=red] node[midway, below, font=\small\bfseries, color=red] {Afterpulsing} (e);

    \end{tikzpicture}
    
    \caption{Schematic representation of the 3-level SPAD carrier dynamics. The solid lines represent standard relaxation and trapping processes. The red dashed line highlights the afterpulsing mechanism.}
    \label{fig:spad_transitions}
\end{figure}
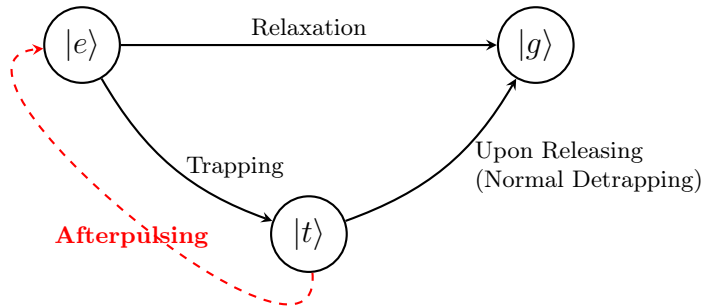

\begin{figure}[H]
    \centering
    \begin{subfigure}[b]{0.49\textwidth}
        \includegraphics[width=\linewidth]{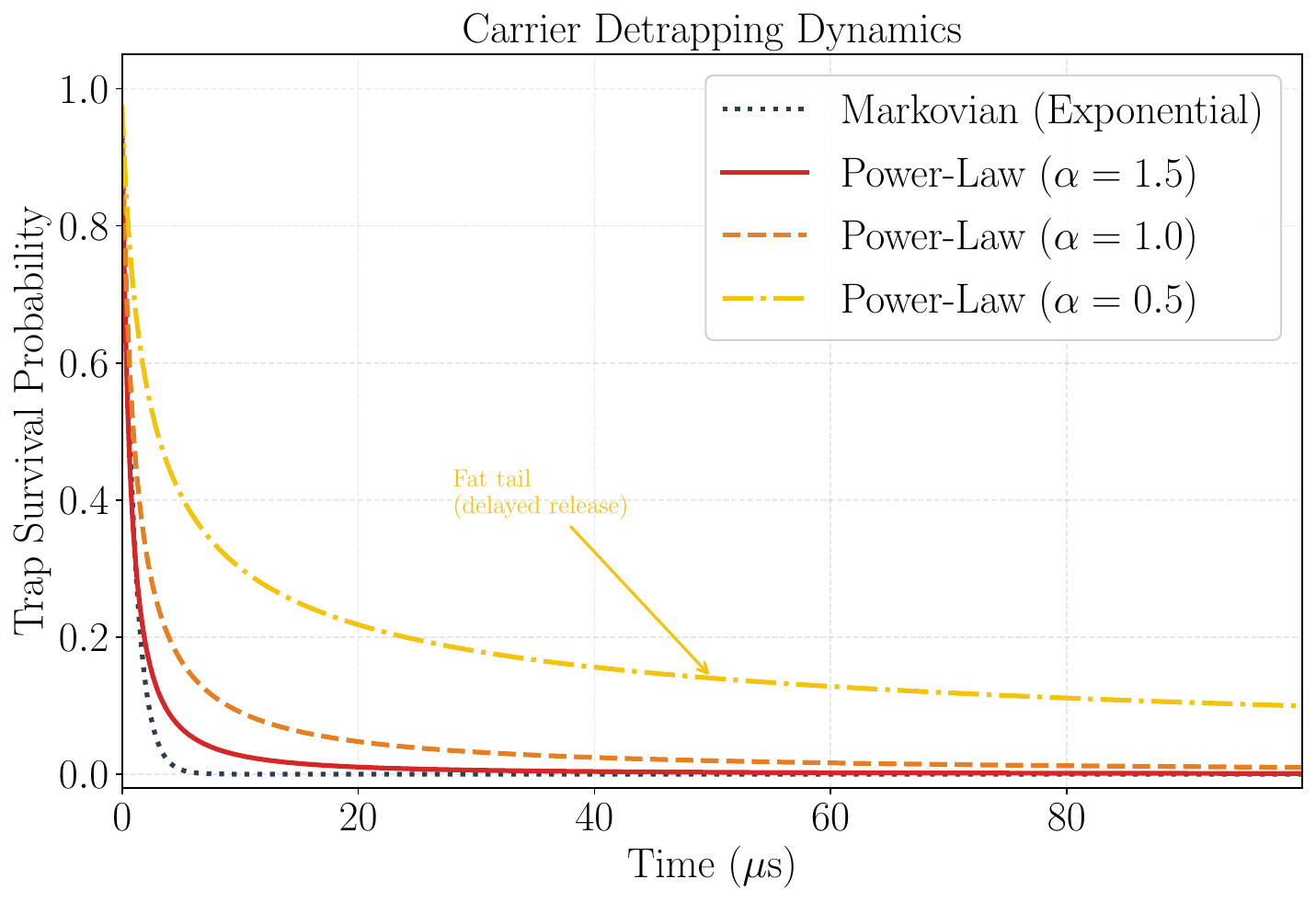}
        \caption{Trap survival probability $P(\ket{t})$ with \ elapsed time for
    Markovian (dotted-black) and power-law models ($\alpha = 1.5$, $1.0$, $0.5$).}
        \label{fig:trap}
    \end{subfigure}
    \hfill 
    \begin{subfigure}[b]{0.49\textwidth}
        \includegraphics[width=\linewidth]{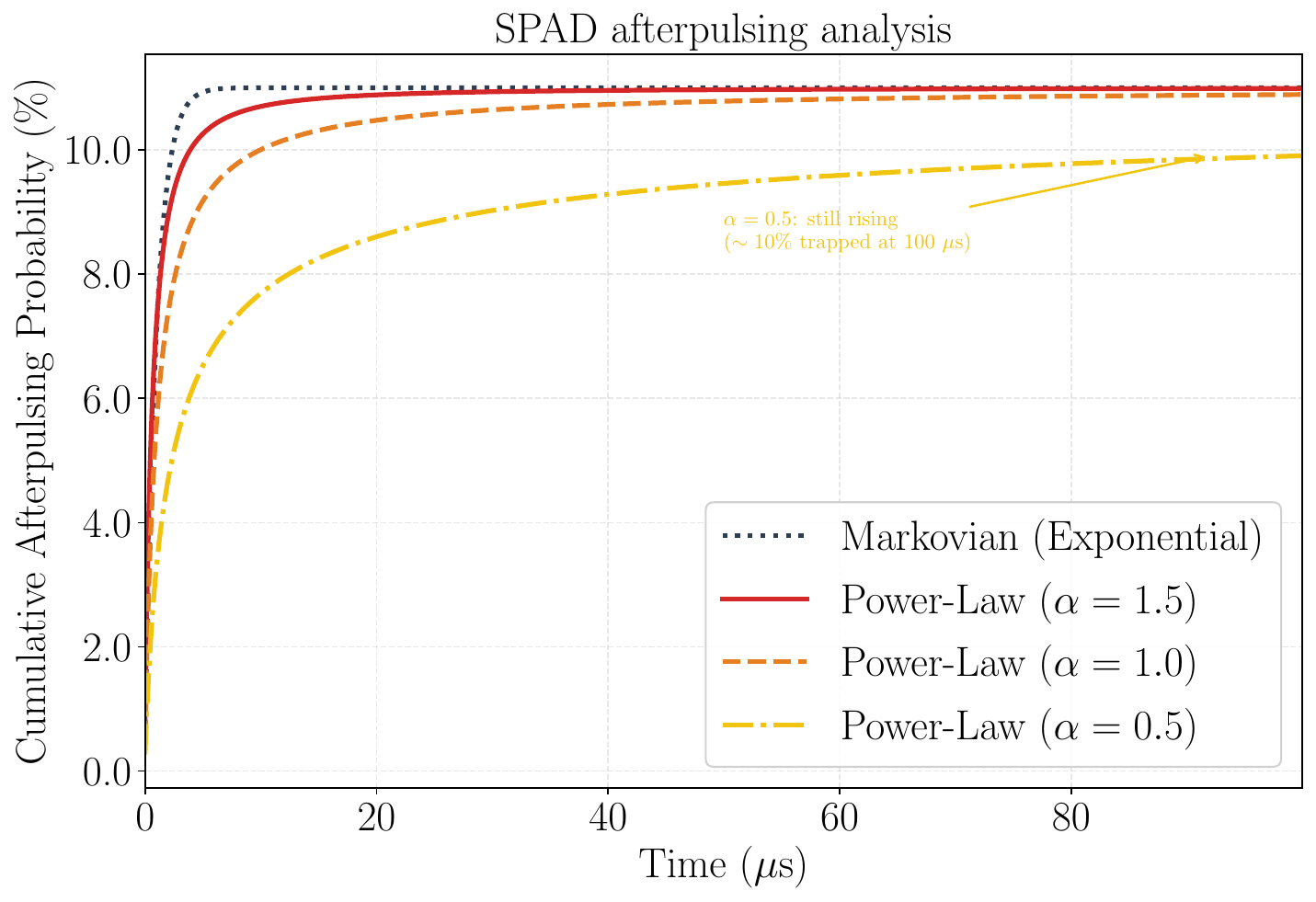}
        \caption{Cumulative afterpulse probability $P_{\mathrm{AP}}$
    accumulated for Markovian (dotted-black) and power-law models ($\alpha = 1.5$, $1.0$, $0.5$).over the same window.}
        \label{fig:afp}
    \end{subfigure}
    \caption{Quantum simulation of SPAD afterpulsing dynamics at proton fluence
    $\Phi = 10^{10}\,\mathrm{cm}^{-2}$.}
    \label{fig:afterpulsing_time}
\end{figure}

Afterpulsing is a false detection event associated with a previous photon detection event. The afterpulsing mechanism is explained in the flowchart of figure~(\ref{fig:spad_transitions}). This figure shows four distinct events that could occur when the SPAD detector absorbs a photon: relaxation, normal detrapping, trapping, and afterpulsing. During the relaxation process, there is a finite probability of a charge carrier becoming trapped in the crystal
defects, denoted by $\ket{t}$ in the figure~(\ref{fig:spad_transitions}). These defects are created either during device fabrication or by
radiation damage from high-energy radiations in the space environment~\cite{disdam, anisimova}. The charge carriers are trapped in the defect for a random amount of time and are then thermally released back into the depletion region, from which they may relax to the ground state $\ket{g}$, or they may trigger a second avalanche, which is a false detection that appears to be a photon detection, but is actually a memory of the previous real photon detection event. Such processes are known as Non-Markovian events, in which the next event depends on the previous one. Hence, to distinguish between real and false detection, it is necessary to simulate afterpulsing. Furthermore, this Quantum simulation allows us to treat trap and excited states as quantum states, and afterpulsing is measured as an expectation value. 
Figure~(\ref{fig:trap}) shows the charge carrier de-trapping probability, while figure~(\ref{fig:afp}) shows the cumulative afterpulsing probability, simulated at the maximum proton fluence
$\Phi = 10^{10}$\,cm$^{-2}$ (equivalent to 24 months in low-Earth
orbit~\cite{anisimova}).
The time window spans 100\,$\mu$s, which is 100 trap lifetimes for
the exponential model (Markovian) \cite{dsouza}.
The aim of this simulation is to compare four trap release models: one Markovian (exponential) and three non-Markovian (power-law with
$\alpha \in \{0.5, 1.0, 1.5\}$) and to show that the shape of the trap release distribution determines the timing and
total count of afterpulses. State evolution is governed by the Jaynes--Cummings Hamiltonian $H_{\mathrm{JCM}}$ with Kraus channels (Eqs.~\ref{eq:kav}--\ref{eq:kstay}) and a passive-quenching reset channel ($\tau_{\mathrm{dead}} = 1.5\,\mu\mathrm{s}$). 

In the figure~(\ref{fig:trap}), the Markovian exponential model is denoted by the black dotted curve, which decays the fastest. It falls to near-zero by approximately $10\,\mu$s. This is physically correct for a device with a single trap species. However, the radiation damage introduces deeper traps. The power-law models are used to observe the charge carrier detrapping  with the the power-law coefficient $\alpha = 1.5, 1.0, 0.5$. It is observed that probability of the carrier de-trapping for $\alpha = 1.5\ \text{and}\ 1.0$ (red and orange curves) decay slowly, with noticeable tails extending beyond $10\,\mu$s. This happens because the power-law distribution includes a range of trap depths. At early times, carriers in shallow traps escape quickly, giving a rapid initial fall. Charge carriers in deeper traps release much more slowly, producing a long tail that the exponential model completely misses. The heavy-tailed model with $\alpha = 0.5$ (yellow curve) shows even at $t = 100\,\mu$s, approximately 10\% of the carrier population is still trapped.
The curve has barely flattened by the end of the simulation window and such a behavior is called a fat tail, and is observed in non-Markovian processes. This is the key physical difference between the Markovian and non-Markovian models. In the Markovian model, every carrier has the same release probability per unit time regardless of its history. In the non-Markovian model, the release probability
$p_\text{release}(t)$ from Eq.~\eqref{eq:prelease_pl} decreases
with time, and the longer the carrier has been trapped, the less likely
it is to escape in the next time step.

The right plot shows the consequence of the left plot, which is the afterpulsing upon releasing from trap levels. The Markovian black curve rises steeply in the first few
microseconds and then saturates cleanly at $P_\text{AP} = 11\%$ by
$t \approx 10\,\mu$s.
This means that with a dead time of $\tau_d = 1.5\,\mu$s, the exponential model predicts that the vast majority of afterpulses
occur within the dead time window and are therefore suppressed. The power-law curves rise more slowly and take much longer
to saturate. The $\alpha = 1.5$ and $\alpha = 1.0$ curves still reach the same
total of $\sim$$11\%$ by the end of the window, in some tens of microseconds rather than a few. This means a significant fraction of their afterpulses occur after the $1.5\,\mu$s dead time has expired, and are counted as a false photon detection. The $\alpha = 0.5$ curve is still not converging and rising at $t = 100\,\mu$s, which implies the afterpulses from deeply trapped carriers continue to arrive for
timescales orders of magnitude longer than the dead time.

A dead time of $\tau_d = 1.5\,\mu$s suppresses nearly all
afterpulses under the exponential model (the trap is $\sim$77\%
emptied by $\tau_d$ for $T_\text{trap} = 1\,\mu$s). Under the $\alpha = 0.5$ power-law model, approximately 50\% of
carriers are still trapped at $t = \tau_d$, meaning half of all
Potential afterpulses occur after the detector resets and
are therefore counted as photon detections.
This directly degrades the quantum bit error rate (QBER) of a QKD
system.

\section{Summary}
In this work, a quantum simulation of a silicon single-photon avalanche diode (SPAD) is carried out. To capture the photon-SPAD interaction at the atomic level using quantum computing, the photon and SPAD are modeled as quantum systems. The projection operators are used to obtain the Hamiltonian matrix for a three-level SPAD detector, while second quantization and Fock-space truncation ensure a finite number of photons in the system. The Jaynes-Cummings (JC) model is used to simulate their interaction. Furthermore, to simulate the JC Hamiltonian on a quantum simulator, the Hamiltonian is mapped to Pauli strings. The Pauli string Hamiltonian thus obtained is evolved in time by using the Lie-Trotter formula on Qiskit. Furthermore, to account for the open-system dynamics of the Photon-SPAD system, the Lindblad master equation and Qiskit's gate-based noise models, such as thermal relaxation operators and Kraus channels, are used.

The efficiency of the SPAD detector is visualized as Rabi oscillations in both an ideal and a damping medium. It is found that photon damping has two effects on SPAD efficiency: it reduces the peak efficiency, and it degrades the timing jitter. This simulation provides a direct, quantitative prediction of SPAD efficiency degradation due to photon dissipation. By varying $\kappa$, the model can map efficiency as a function of the dissipation rate for any given material or cavity geometry. Furthermore, the thermal dark count probability is simulated at various temperature ranges using the thermal relaxation error channel. To find the excited state population parameter of the DCR simulation, the Fermi-Dirac statistics is utilized. Both approaches for simulating thermal dark counts produce the same behavior. Furthermore, the simulation framework allows the dark count rate to be computed for different types of SPAD detectors by substituting appropriate parameters, such as the activation energy $\Delta E$ and relaxation time. 

The afterpulsing of SPAD is modeled by Kraus noise channels. A power-law expansion is used to model the trap-release time, as it is strongly influenced by external radiation. This simulation provides important insights into dead-time design to reduce afterpulsing counts and improve QKD efficiency. Afterpulses that occur after the nominal dead time are correlated with previous detections. An
eavesdropper aware of this correlation could exploit it. The power-law model reveals a much longer
correlation window than the exponential model, which tightens the security analysis. For radiation-damaged SPADs in low-Earth orbit, the quenching dead time must be set based on the power-law trap release distribution, not the exponential approximation. This quantum simulation provides a quantitative prediction of the required dead time for a given
fluence and power law coefficient, $\alpha$. Furthermore, the framework is
material-agnostic. By substituting DFT-computed trap energies for other materials (such as hBN, WSe$_2$) into the power-law parameters, the
same simulation can predict their afterpulsing behavior. Thermal annealing at $+50$ to
$+100\,^{\circ}$C reduces displacement-damage trap density. In a non-Markovian picture, annealing is most effective when it
specifically removes the shallow traps that sustain the fat tail
(low $\alpha$ carriers).  The simulation can be run before and after an annealing
event by updating $\tau_0$ and $\alpha$, directly predicting the success of annealing.
\section*{Acknowledgments}

The authors acknowledge MNIT Jaipur and the Ministry of Education, Government of India, for providing the support to execute this research work.
\newpage
\appendix
\renewcommand{\thesection}{Appendix \Alph{section}}

\section{Photon-SPAD interaction Hamiltonian} \label{appA}
By performing the tensor products block-by-block of the interaction Hamiltonian of Eq.~(\ref{rwa}), the first and second terms are expanded as follows.
\begin{equation}  \label{int1}
    a^\dagger \otimes \sigma_{ge} = \begin{pmatrix} 
    \mathcal{O} & \mathcal{O} & \mathcal{O} & \mathcal{O} \\ 
    \sigma_{ge} & \mathcal{O} & \mathcal{O} & \mathcal{O} \\ 
    \mathcal{O} & \sqrt{2}\sigma_{ge} & \mathcal{O} & \mathcal{O} \\ 
    \mathcal{O} & \mathcal{O} & \sqrt{3}\sigma_{ge} & \mathcal{O} 
    \end{pmatrix}
\end{equation}

\begin{equation}
    a \otimes \sigma_{eg} = \begin{pmatrix} 
    \mathcal{O} & \sigma_{eg} & \mathcal{O} & \mathcal{O} \\ 
    \mathcal{O} & \mathcal{O} & \sqrt{2}\sigma_{eg} &\mathcal{O} \\ 
   \mathcal{O} & \mathcal{O} &\mathcal{O} & \sqrt{3}\sigma_{eg} \\ 
   \mathcal{O} & \mathcal{O} & \mathcal{O} & \mathcal{O}
    \end{pmatrix}
\end{equation}
where $\mathcal{O}$ represents a $4 \times 4$ null matrix, while $\sigma_{eg}\ \&\ \sigma_{ge}$ are the $4\times 4$ SPAD raising and lowing operators. Upon expressing the matrices $\mathcal{O}$, $\sigma_{ge}$ and $\sigma_{eg}$, the full $16\times 16$ matrix is expressed as,
\begin{equation} \label{16intmatrix}
H_{\mathrm{int}} = \hbar g
\left(
\begin{array}{cccc|cccc|cccc|cccc}
0 & 0 & 0 & 0 & 0 & 0 & 0 & 0 & 0 & 0 & 0 & 0 & 0 & 0 & 0 & 0 \\
0 & 0 & 0 & 0 & 1 & 0 & 0 & 0 & 0 & 0 & 0 & 0 & 0 & 0 & 0 & 0 \\
0 & 0 & 0 & 0 & 0 & 0 & 0 & 0 & 0 & 0 & 0 & 0 & 0 & 0 & 0 & 0 \\
0 & 0 & 0 & 0 & 0 & 0 & 0 & 0 & 0 & 0 & 0 & 0 & 0 & 0 & 0 & 0 \\
\hline
0 & 1 & 0 & 0 & 0 & 0 & 0 & 0 & 0 & 0 & 0 & 0 & 0 & 0 & 0 & 0 \\
0 & 0 & 0 & 0 & 0 & 0 & 0 & 0 & \sqrt{2} & 0 & 0 & 0 & 0 & 0 & 0 & 0 \\
0 & 0 & 0 & 0 & 0 & 0 & 0 & 0 & 0 & 0 & 0 & 0 & 0 & 0 & 0 & 0 \\
0 & 0 & 0 & 0 & 0 & 0 & 0 & 0 & 0 & 0 & 0 & 0 & 0 & 0 & 0 & 0 \\
\hline
0 & 0 & 0 & 0 & 0 & \sqrt{2} & 0 & 0 & 0 & 0 & 0 & 0 & 0 & 0 & 0 & 0 \\
0 & 0 & 0 & 0 & 0 & 0 & 0 & 0 & 0 & 0 & 0 & 0 & \sqrt{3} & 0 & 0 & 0 \\
0 & 0 & 0 & 0 & 0 & 0 & 0 & 0 & 0 & 0 & 0 & 0 & 0 & 0 & 0 & 0 \\
0 & 0 & 0 & 0 & 0 & 0 & 0 & 0 & 0 & 0 & 0 & 0 & 0 & 0 & 0 & 0 \\
\hline
0 & 0 & 0 & 0 & 0 & 0 & 0 & 0 & 0 & \sqrt{3} & 0 & 0 & 0 & 0 & 0 & 0 \\
0 & 0 & 0 & 0 & 0 & 0 & 0 & 0 & 0 & 0 & 0 & 0 & 0 & 0 & 0 & 0 \\
0 & 0 & 0 & 0 & 0 & 0 & 0 & 0 & 0 & 0 & 0 & 0 & 0 & 0 & 0 & 0 \\
0 & 0 & 0 & 0 & 0 & 0 & 0 & 0 & 0 & 0 & 0 & 0 & 0 & 0 & 0 & 0
\end{array}
\right)
\end{equation}

\section{Initial and observable states of SPAD quantum simulation} \label{appB}
Many possible initial quantum states and their respective Qiskit basis (binary-encoded) states are listed in Table \ref{tab:Initial}. 
\begin{table}[H]
    \centering
    \renewcommand{\arraystretch}{1.1}
    \setlength{\tabcolsep}{8pt}       
    \caption{Mapping of Physical SPAD--Photon States to Binary and Qiskit Basis}
    \label{tab:Initial}
    
    \begin{tabular}{ l l c } 
        \toprule
        
        \rowcolor{blue!10} 
        \textbf{Physical state} & \textbf{Binary} & \textbf{Qiskit} \\
        \midrule
        Single photon + SPAD armed $|g\rangle $ & $|01\rangle_p|00\rangle_s$ & $|0100\rangle$ \\
        Two photons + SPAD armed  & $|10\rangle_p|00\rangle_s$ & $|1000\rangle$ \\
        Three photons + SPAD armed & $|11\rangle_p|00\rangle_s$ & $|1100\rangle$ \\
        No photon + SPAD excited $|e\rangle$  & $|00\rangle_p|01\rangle_s$ & $|0001\rangle$ \\
        No photon + SPAD in trap state & $|00\rangle_p|10\rangle_s$ & $|0010\rangle$ \\
        one photon + SPAD excited & $|01\rangle_p|01\rangle_s$ & $|0101\rangle$ \\ 
        \bottomrule
    \end{tabular}
\end{table}
Furthermore, after defining the initial state, the observable state is defined. According to Eq. (\ref{jcref}), the observable state is defined such that the excitation number remains conserved. However, when the system is open, some transitions that are not ideally possible occur. These transitions are governed by noise channels such as thermal noise and silicon lattice defects. The initial and observable states, with their respective physical meanings used in the quantum simulation, are defined in the table \ref{tab:obs}.
\begin{table}[H]
    \centering
    \renewcommand{\arraystretch}{1.3}
    \setlength{\tabcolsep}{3pt}
    \caption{Mapping of Physical SPAD--Photon States to Binary and Qiskit Basis}
    \label{tab:obs}
    
    \resizebox{\textwidth}{!}{%
    \begin{tabular}{ l l p{9cm} }
        \toprule
        \rowcolor{blue!10} 
        \textbf{Initial state} & \textbf{Possible observable state} & \textbf{Physical Meaning} \\
        \midrule
        $|0100\rangle$ & $\ket{0001}$ & System is initialized with one photon and SPAD in $\ket{g}$ state, and the probability to find SPAD to be in the excited state is observed \\

        $|0000\rangle$ & $\ket{0001}$ & The system is initialized in the vacuum state, and the probability to be found in the excited state (the dark counts) is observed \\

        $|0100\rangle$ & $\ket{0010}$ & System is initialized with one photon and SPAD in $\ket{g}$ state, and the probability to find SPAD to be in the trap state is observed \\

        $|0010\rangle$ & $\ket{0100}$ & System is initialized with the SPAD in $\ket{t}$ state, and the probability of afterpulsing is observed \\

        $|0100\rangle$ & $\ket{0000}$ & System is initialized with the one photon and SPAD in $\ket{g}$ state, and the probability of photon dissipation is observed \\
        
        \bottomrule
    \end{tabular}%
    }
\end{table}
\section{Time evolution of Photon-SPAD system} \label{appC}
In closed quantum systems, this is governed by the unitary time evolution operator, which is expressed in Eq.~(\ref{timeevolve}).
\begin{equation} \label{timeevolve}
    U(t) = e^{-i H_{\mathrm{JCM}} t / \hbar}
\end{equation}
where $H_{\mathrm{JCM}}$ is the Jaynes-Cummings Hamiltonian. To execute this continuous evolution on a quantum computer, the unitary operator is decomposed into a sequence of discrete quantum gates. For this purpose, however, the total Hamiltonian is decomposed into the Pauli strings, $H_{\mathrm{JCM}} = \sum_{i} c_i P_i$, as described in section \ref{3.1}. In standard algebra, the exponential of a sum is equal to the product of the individual exponentials. However, because the Pauli matrices generally do not commute, i.e., $[P_i, P_j] \neq 0$, this rule breaks down for quantum operators.
\begin{equation}
    e^{A+B} \neq e^A e^B
\end{equation}
Consequently, the total Hamiltonian can not be simulated simply by applying the quantum gates. In this case, the Eq.~(\ref{timeevolve}) above is false.

\section{Trotterized quantum circuit} \label{appD}
The two Pauli X gates initialize the quantum circuit in state $\ket{1100}$ in standard little-Endian notation, which is followed by the Trotterization circuit. The decomposed trotterization circuit for one trotter step is shown below.
\begin{figure}[H]
     \centering
     \includegraphics[width=1\linewidth]{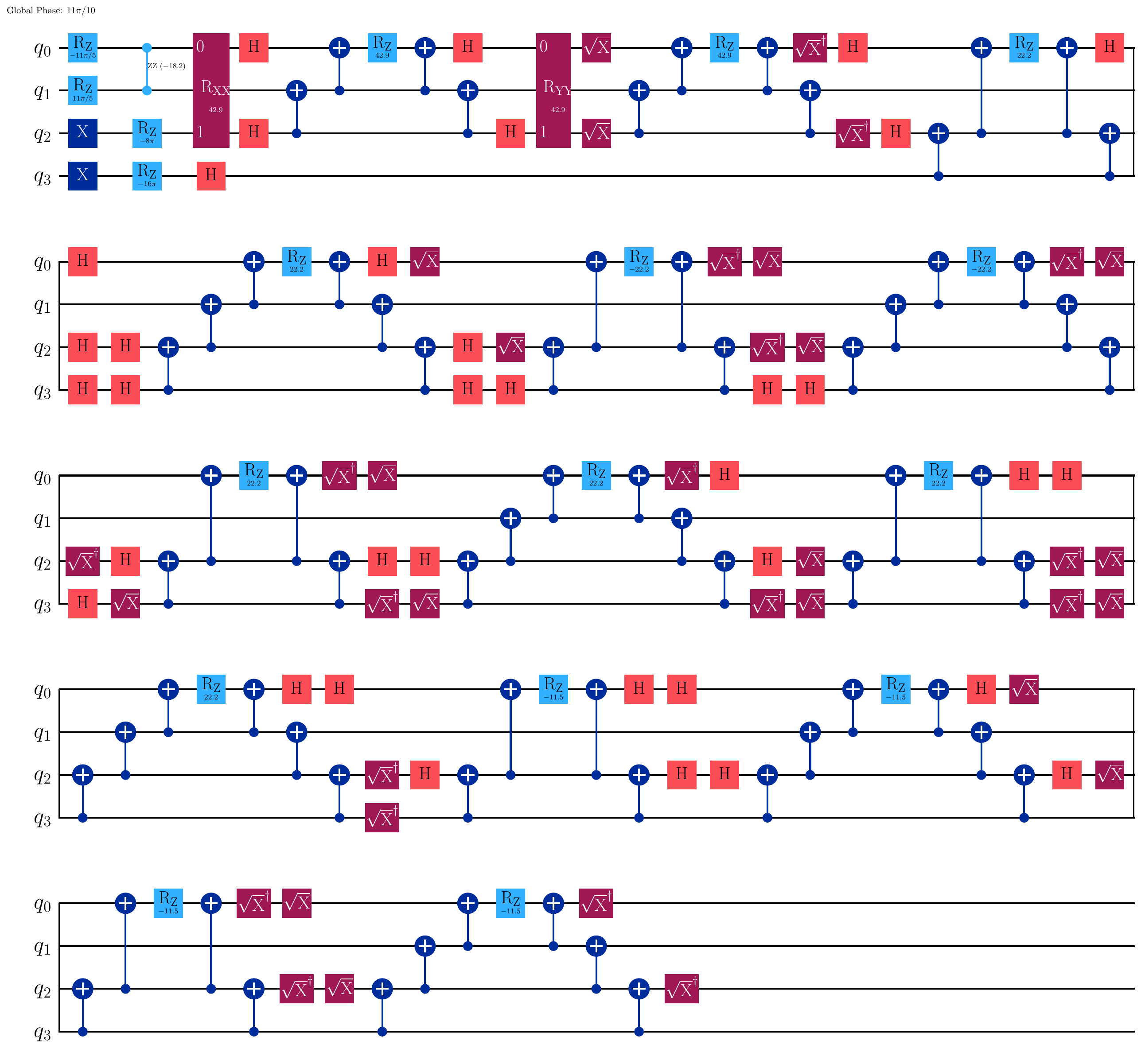}
     \caption{Decomposed Trotter circuit for the first step, using \texttt{quantumcircuit.decompose()}.}
     \label{fig:decomposed}
 \end{figure}
\section{Trace Preservation of the Afterpulsing Kraus Channel}
\label{app:trace_preservation}

This appendix derives the trace-preservation condition used to validate the afterpulsing
Kraus channel of Eqs.~(\ref{eq:kav})--(\ref{eq:kstay}). A quantum channel $\mathcal{E}$ defined by Kraus operators $\{K_i\}$ acts on a density
matrix as
\begin{equation}
    \mathcal{E}(\rho) = \sum_i K_i \, \rho \, K_i^\dagger .
    \label{eq:app_kraus_channel}
\end{equation}
For $\mathcal{E}$ to describe a physical process, total probability must be conserved,
i.e.\ $\operatorname{Tr}[\mathcal{E}(\rho)] = \operatorname{Tr}[\rho]$ for every $\rho$.
Using the cyclic property of the trace, $\operatorname{Tr}[ABC] = \operatorname{Tr}[CAB]$,

\begin{align}
    \operatorname{Tr}\!\big[\mathcal{E}(\rho)\big]
        &= \operatorname{Tr}\!\left[\sum_i K_i \rho K_i^\dagger\right]
         = \sum_i \operatorname{Tr}\!\left[K_i^\dagger K_i \rho\right] \\
        &= \operatorname{Tr}\!\left[\left(\sum_i K_i^\dagger K_i\right)\rho\right].
\end{align}

This equals $\operatorname{Tr}[\rho]$ for arbitrary $\rho$ if and only if

\begin{equation}
    \sum_i K_i^\dagger K_i = \mathbf{I}.
    \label{eq:app_completeness}
\end{equation}

Eq.~(\ref{eq:app_completeness}) is therefore the necessary and sufficient condition for
a set of Kraus operators to define a physical, trace-preserving channel. Hence, expanding the RHS part of Eqs.~(\ref{eq:kav})--(\ref{eq:kstay}), 

\begin{align}
    \sum_i K_i^\dagger K_i
        &= \Big[\,P_{\mathrm{release}}\,P_{\mathrm{ap}} + P_{\mathrm{release}}(1-P_{\mathrm{ap}}) + (1-P_{\mathrm{release}})\,\Big]\ket{t}\bra{t}
           + \sum_{i\neq t}\ket{i}\bra{i} \label{eq:app_step1}\\
        &= \Big[\,P_{\mathrm{release}}\underbrace{(P_{\mathrm{ap}} + 1 - P_{\mathrm{ap}})}_{=\,1}
              + (1-P_{\mathrm{release}})\,\Big]\ket{t}\bra{t}
           + \sum_{i\neq t}\ket{i}\bra{i} \label{eq:app_step2}\\
        &= \underbrace{\big[\,P_{\mathrm{release}} + 1 - P_{\mathrm{release}}\,\big]}_{=\,1}\ket{t}\bra{t}
           + \sum_{i\neq t}\ket{i}\bra{i} \label{eq:app_step3}\\
        &= \ket{t}\bra{t} + \sum_{i\neq t}\ket{i}\bra{i}
         = \mathbf{I}.
    \label{eq:app_final}
\end{align}
where $i \in \{g, e\}$. Hence, the condition described in Eq.~(\ref{eq:app_completeness}) is fulfilled.

\section{Simulation parameters} \label{appF}
Table~\ref{tab:params} lists all rates used in the simulation together with
their physical meaning and source.
 \begin{table}[H]
    \centering
    \small
    \renewcommand{\arraystretch}{1.5}
    \setlength{\tabcolsep}{6pt}
    \caption{}
    \label{tab:params}
    \begin{tabular}{p{3cm} p{6.5cm} p{3.9cm}}
        \toprule
        \rowcolor{blue!10}
        \textbf{Parameter} & \textbf{Physical meaning} &
        \textbf{Value and source} \\
        \midrule
        Photon-SPAD coupling constant $g$ &
        Photon--SPAD interaction strength in the Jaynes-Cummings
        Hamiltonian. &
        $0.1$ in simulation units, whose physical value is $\approx 7.48 \times 10^8$ \cite{meher} \\
        Number of trotter steps & The number of times one trotter step shown in figures~(\ref{fig:trottercirc}) or (\ref{fig:decomposed}) repeats after initial quantum state. Used to increase accuracy of the quantum simulation. & 200\\
        Photon dissipation rate $\kappa$ &
        The rate at which a cavity photon is absorbed or scattered by the medium,
        reducing detection efficiency. &
        $\frac{g}{2}=0.05$ in simulation units \cite{meher} \\
 
        Rate of Thermal excitation $\gamma_\text{th}$ &
        Rate of thermally induced dark counts $\ket{g}\to\ket{e}$, computed
        from the Arrhenius equation \ref{eq:arrhenius}. &
        Varied according to the Arrhenius principle at the temperature rate,
        $T = 173,\ 273\ \&\ 303\ K$~\cite{anisimova, IEEE, Sze2006} \\
 
        Simulation gate time $t_\text{gate}$/ Simulation unit used in thermal dark count simulation &
        The physical duration of one circuit time step, used to convert
        physical rate $\gamma_{th}$ (s$^{-1}$) to dimensionless simulation units, and in qiskit's \texttt{thermal\_relaxation\_error} channel.&
        $50$\,ns \\
 
        Reference proton fluence $\phi_0$ &
        Radiation dose at which the reference afterpulsing probability $P_{ap, 0}$ is measured in Eq.~(\ref{eq:pap}) &
        $1\times10^{9}$\,p\,cm$^{-2}$~\cite{anisimova} \\
 
        Reference Afterpulsing probability $P_{ap.0}$ at reference fluence &
        Probability that a trapped carrier re-triggers an avalanche,
        measured at $\phi_0$. &
        $1.1\,\%$~\cite{anisimova} \\

        Trap state lifetime $\tau_o$& Time after which charge carrier leave the trap state. &
        $10^{-6} s$ \cite{disdam} \\
 
        \bottomrule
    \end{tabular}
\end{table}

\section{Derivation of simulation unit to calculate timing jitter} \label{appE}
A previous study \cite{meher} regarding the strong coupling between a semiconductor double quantum dot and a superconducting
microwave cavity, resolving a vacuum Rabi mode splitting of
$2g/2\pi = 238$\,MHz. This gives a coupling strength of,
\begin{equation}
    g_\text{phys} = \frac{2g}{2} = \frac{2\pi \times 238 \times 10^6}{2}
    \approx 7.48 \times 10^8 \ \text{rad\,s}^{-1}
    \label{eq:g_phys}
\end{equation}
In this simulation, $g = 0.1$ represents this physical coupling. Hence, the simulation time unit can be derived as,
\begin{equation}
    t_\text{unit}
    = \frac{g_\text{sim}}{g_\text{phys}}
    = \frac{0.1}{7.48 \times 10^8}
    \approx 1.34 \times 10^{-10} \ \text{s}
    = 0.134 \ \text{ns}
    \label{eq:t_unit}
\end{equation}

Substituting into Eq.~\eqref{eq:fwhm_sim}, the value of timing jitter is obtained.
\begin{equation}
    \Delta t_\text{jitter}
    = 6.62 \times 0.134 \ \text{ns}
    \approx \mathbf{0.887 \ \text{ns} = 887 \ \text{ps}}
    \label{eq:jitter_physical}
\end{equation}
In a similar way, the timing jitter of an open-system SPAD detector can also be obtained.
\bibliographystyle{unsrt}


\end{document}